\documentclass[10pt,a4paper,twocolumn]{article}
\usepackage{styles/preprint}
\newcommand{\ManuscriptTitle}{A Unified AI Framework for Rapid and Accurate Waveform Generation of Compact Binary Coalescences}
\newcommand{\AuthorNames}{Mengfei Sun\textsuperscript{1,2}, Jin Li\textsuperscript{1,2,3,*}, Yuanhong Zhong\textsuperscript{7}, Jie Wu\textsuperscript{1,2}, Borui Wang\textsuperscript{4,5}, Xianghe Ma\textsuperscript{4}, Minghui Zhang\textsuperscript{6}}
\newcommand{\AuthorAffiliationsLeft}{%
\textsuperscript{1}Department of Physics, Chongqing University, Chongqing 401331, People's Republic of China.\par
\textsuperscript{2}Chongqing Key Laboratory for Strongly Coupled Physics, Chongqing University, Chongqing 401331, People's Republic of China.\par
\textsuperscript{3}Institute of Advanced Interdisciplinary Studies, Chongqing University, Chongqing 401331, China.}
\newcommand{\AuthorAffiliationsRight}{%
\textsuperscript{4}Department of Earth and Space Sciences, Southern University of Science and Technology, Shenzhen 518055, People's Republic of China.\par
\textsuperscript{5}School of Astronomy and Space Science, Nanjing University, Nanjing 210023, People's Republic of China.\par
\textsuperscript{6}Department of Physics, Southern University of Science and Technology, Shenzhen 518055, People's Republic of China.\par
\textsuperscript{7}School of Microelectronics and Communication Engineering, Chongqing University, Chongqing 401331, People's Republic of China.\par
\textsuperscript{*}Corresponding author email \href{mailto:cqujinli1983@cqu.edu.cn}{cqujinli1983@cqu.edu.cn}.}
\newcommand{\AuthorAffiliations}{\AuthorAffiliationsLeft\par\AuthorAffiliationsRight}

\hypersetup{
  pdftitle={\ManuscriptTitle},
  pdfauthor={Mengfei Sun, Jin Li, Yuanhong Zhong, Jie Wu, Borui Wang, Xianghe Ma, Minghui Zhang},
  pdfsubject={Compact binary waveform generation with URSA-GW}
}
\begin{document}
\MakePreprintTitle
\section{Introduction}
Gravitational wave (GW) observations have opened a new window for testing strong field gravity, investigating the formation and evolution of compact objects, probing matter above nuclear saturation density, and measuring the expansion history of the Universe~\cite{Sathyaprakash2009,Abbott2018EOS,Abbott2017StandardSiren,Hannam2022Precession}. GW150914, GW170817, and GW200115 are representative events of binary black hole (BBH), binary neutron star (BNS), and neutron star black hole (NSBH) mergers, respectively~\cite{GW150914,GW170817,Abbott2021NSBH,GWTC3}. Extracting physical information from these observations requires waveform models that accurately describe binary dynamics and matter effects. Next generation detectors will observe more distant compact binaries and substantially enlarge the observable source population. They will also track inspirals from lower frequencies and record longer gravitational waveforms~\cite{Bailes2021FutureGW}. Waveform models must be evaluated repeatedly in simulations, event searches, and parameter inference, so the cost of individual evaluations accumulates as the numbers of simulated waveforms, search templates, and inference samples increase. With larger observational datasets, expensive waveform calculations will limit parameter space coverage in simulations and searches, increase the time required for event parameter inference~\cite{HuVeitch2025Cost}, and constrain systematic studies of compact binary populations.

Compact binary waveforms reflect orbital evolution, spin dynamics, and the material properties of the component objects. The inspiral, merger, and ringdown of BBHs are described by the vacuum field equations of general relativity. Effective one body (EOB) and frequency domain phenomenological models have incorporated spin precession and higher order radiation modes into waveform modeling~\cite{Buonanno1999,Pompili2023,SEOBNRv5PHM,IMRPhenomXPHM}. For BNS and NSBH systems, the neutron star tidal response further modifies the inspiral waveform and carries information about internal structure and the equation of state~\cite{Flanagan2008Tidal,Hinderer2008Love,Hinderer2016DynamicTides,NRTidalv3}. In NSBH systems, the mass ratio, black hole spin, and neutron star compactness jointly affect tidal disruption, changing the radiation at high frequencies and the spectral cutoff near merger~\cite{TEOBBHNS}. If a BNS merger leaves a neutron star remnant, its oscillations produce gravitational waves whose characteristic frequencies are closely related to the remnant structure and equation of state, providing information about matter beyond inspiral tides~\cite{Bauswein2012Postmerger,Bernuzzi2015Spectrum,Stergioulas2011Modes,NRPM2019}. Effective one body, phenomenological, and postmerger models developed for these processes describe the waveforms of different binary types and provide a physical basis for fast surrogate models.

Reduced order modeling and neural networks provide effective ways to accelerate waveform generation based on physical waveform models. Reduced bases and empirical interpolation exploit correlations between waveforms to represent them with a small number of basis functions and sampling points~\cite{Field2011,Field2014}. These methods have been applied to numerical relativity~\cite{Blackman2015,Blackman2017}, BBH waveforms and extensions including higher order modes~\cite{Purrer2014,Varma2019Hybrid}, and BNS waveforms with tidal effects~\cite{Lackey2019}. Related reduced order surrogate models have also been extended to long duration eccentric waveforms~\cite{Maurya2026}. Studies of waveform reconstruction for massive BBHs observed by the Laser Interferometer Space Antenna (LISA) have explored low dimensional nonlinear and adaptive time--frequency representations~\cite{Leroy2024}. Neural networks learn mappings from binary parameters to waveforms or their low dimensional coefficients, reducing repeated evaluations of the original model during generation~\cite{Chua2019ROM,Liao2021CAE}. Existing studies cover BBHs, BNS tidal inspirals, and precessing waveforms with multiple modes~\cite{ANNSur2021,Grimbergen2024HM,Freitas2025DANSur,mlgwBNS2023,Sun2025cAE,Whittall2026}, and extend to extreme mass ratio inspirals and eccentric systems~\cite{Chua2021,Shi2025Eccentric,Shi2025HigherModes}. These methods have enabled rapid waveform generation in complex cases involving precession, higher order modes, and eccentricity, but their waveform representations and parameter mappings are mainly constructed separately for specific binary types or reference models. For compact binaries with common orbital dynamics and different matter effects, an open question is whether shared modeling can reduce reliance on separate generation models.

Combining physical modeling with machine learning~\cite{Carleo2019,Karniadakis2021} can exploit dynamical features shared by different binary types. Known inspiral laws describe the leading waveform trends and allow different binary types to be represented on a common dynamical basis. Neural networks learn amplitude and phase corrections of the reference waveforms relative to a physical baseline. For common dynamical components controlled by parameters such as the mass ratio and spins, the parameters of this mapping can be learned jointly across binary types. Tidal disruption and BNS postmerger radiation are treated according to their respective physical mechanisms and connected to the shared description of the dynamics.

We present URSA-GW, the Unified Residual Spectral Approximant for Gravitational Waves. Compared with generation methods constructed separately for specific binary types, URSA-GW brings BBH, BNS, and NSBH waveforms into a unified framework through a common analytic residual spectral representation and a shared parameter mapping, while modules specific to each binary type describe their physical differences. The network predicts a small set of spectral coefficients and postmerger physical parameters, which are combined with an analytic baseline and physical decoding to recover the waveforms. This approach integrates generation across binary types, the description of complex physical effects, and rapid evaluation within one procedure, while retaining module interfaces for extending the frequency coverage and physical effects. The model retains spin precession, higher order radiation modes, and neutron star matter effects present in the reference waveforms of the respective binary types, including BNS postmerger radiation. Mean mismatches relative to the reference waveforms range from $5.89\times10^{-4}$ to $5.34\times10^{-2}$ across the three binary types, and batch generation on a graphics processing unit (GPU) achieves a maximum speedup of approximately $1.8\times10^{3}$ over reference calculations on 64 central processing unit (CPU) cores. This design provides a unified, efficient waveform generation method for large scale waveform simulations, event searches, parameter inference, and astrophysical research.

\section{Results}
\subsection{Unified waveform generation across binary types}
\begin{figure*}[tbp]
\centering
\resizebox{\textwidth}{!}{
\begingroup
\definecolor{ConceptBBH}{HTML}{C58B25}
\definecolor{ConceptBNS}{HTML}{3678B5}
\definecolor{ConceptNSBH}{HTML}{168C87}
\definecolor{ConceptInk}{HTML}{203247}
\definecolor{ConceptLine}{HTML}{CDD7E1}
\begin{tikzpicture}[x=1cm,y=1cm,font=\sffamily\fontsize{8.5}{10}\selectfont,
  text=ConceptInk,line cap=round,line join=round,
  flow/.style={-{Stealth[length=2mm,width=1.4mm]},line width=.7pt,draw=ConceptInk!65},
  head/.style={font=\sffamily\bfseries\fontsize{10.5}{12}\selectfont},
  note/.style={font=\sffamily\fontsize{7.5}{9}\selectfont},
  panel/.style={draw=ConceptLine,line width=.55pt,rounded corners=1.5mm,fill=white}]
\path[use as bounding box] (0,3.0) rectangle (17.55,10.05);
\node[anchor=west,head] at (0,9.7) {a};
\node[head] at (1.85,9.7) {Binary types};
\node[head] at (5.95,9.7) {Shared inspiral structure};
\node[head] at (12.9,9.7) {Distinct merger physics};

\foreach \yy/\cc/\ss in {8.55/ConceptBBH/BBH,6.45/ConceptBNS/BNS,4.35/ConceptNSBH/NSBH}{
  \node[anchor=west,font=\sffamily\bfseries\fontsize{11}{12}\selectfont,text=\cc] at (.05,\yy) {\ss};
  \draw[\cc!45,line width=.65pt,rotate around={-14:(2.1,\yy)}] (2.1,\yy) ellipse (.75 and .30);
  \draw[flow,draw=\cc] (2.95,\yy)--(3.35,\yy);
  \draw[flow,draw=\cc] (8.05,\yy)--(8.55,\yy);
}
\foreach \xx/\dy in {1.67/.11,2.53/-.11}{
  \fill[ConceptInk] (\xx,8.55+\dy) circle (.19);
  \draw[ConceptBBH!65,line width=.6pt] (\xx,8.55+\dy) circle (.235);
}
\foreach \xx/\dy in {1.67/.11,2.53/-.11}{
  \filldraw[fill=ConceptBNS!14,draw=ConceptBNS,line width=.85pt] (\xx,6.45+\dy) circle (.19);
  \draw[ConceptBNS!65,line width=.5pt] (\xx,6.45+\dy) ellipse (.085 and .19);
  \draw[ConceptBNS!65,line width=.5pt] (\xx,6.45+\dy) ellipse (.19 and .07);
}
\fill[ConceptInk] (1.67,4.46) circle (.22);
\draw[ConceptNSBH!70,line width=.6pt] (1.67,4.46) circle (.265);
\filldraw[fill=ConceptNSBH!14,draw=ConceptNSBH,line width=.8pt] (2.53,4.24) circle (.16);
\draw[ConceptNSBH!65,line width=.5pt] (2.53,4.24) ellipse (.07 and .16);

\foreach \ss/\cc/\yy/\leftlo/\lefthi/\rightlo/\righthi in {
 BBH/ConceptBBH/8.55/-330/-110/-35/22,
 BNS/ConceptBNS/6.45/-30/-12/1/5,
 NSBH/ConceptNSBH/4.35/-40/-18/-6/3}{
  \foreach \stage/\xx/\tlo/\thi in {inspiral/3.65/\leftlo/\lefthi,merger/8.85/\rightlo/\righthi}{
    \begin{scope}[shift={(\xx,\yy)},x=4.0cm,y=.52cm]
      \draw[\cc!18,line width=.4pt] (0,0)--(1,0);
      \draw[\cc,line width=.8pt] plot file {figures/editable_concept/\ss_\stage.dat};
    \end{scope}
    \draw[ConceptLine,line width=.5pt] (\xx,\yy-.68)--(\xx+4,\yy-.68);
    \draw[ConceptInk!65,line width=.5pt] (\xx,\yy-.64)--(\xx,\yy-.72);
    \draw[ConceptInk!65,line width=.5pt] (\xx+4,\yy-.64)--(\xx+4,\yy-.72);
    \node[note,anchor=north] at (\xx,\yy-.78) {$\tlo$};
    \node[note,anchor=north] at (\xx+4,\yy-.78) {$\thi$};
    \node[note,anchor=north] at (\xx+2,\yy-.78) {$t-t_{\rm peak}$ (ms)};
  }
}

\fill[ConceptInk] (13.65,8.55) circle (.28);
\draw[ConceptBBH!65,line width=.7pt] (13.65,8.55) ellipse (.39 and .30);
\draw[ConceptBBH!30,line width=.6pt] (13.65,8.55) ellipse (.47 and .34);
\node[anchor=west,align=left] at (14.35,8.55) {Merger and\\ringdown};
\filldraw[ConceptBNS,line width=.8pt,fill=ConceptBNS!14] (13.65,6.45) ellipse (.28 and .20);
\draw[ConceptBNS!60,line width=.7pt] (13.65,6.45) ellipse (.36 and .29);
\draw[ConceptBNS!30,line width=.6pt] (13.65,6.45) ellipse (.43 and .36);
\node[anchor=west,align=left] at (14.35,6.45) {Postmerger\\oscillations};
\fill[ConceptInk] (13.48,4.35) circle (.23);
\draw[ConceptNSBH!65,line width=.65pt] (13.48,4.35) circle (.28);
\filldraw[fill=ConceptNSBH!14,draw=ConceptNSBH,line width=.7pt]
 (13.78,4.43) .. controls (14.25,4.78) and (14.35,4.25) .. (13.78,4.43);
\draw[ConceptNSBH!65,line width=.65pt] (13.91,4.55) .. controls (14.20,4.84) and (13.51,4.96) .. (13.22,4.61);
\draw[ConceptNSBH!40,line width=.55pt] (13.91,4.55) .. controls (14.38,4.95) and (13.38,5.02) .. (13.08,4.55);
\node[anchor=west,align=left] at (14.35,4.35) {Tidal effects\\and disruption};

\end{tikzpicture}
\endgroup}
\par\vspace{5pt}
\resizebox{\textwidth}{!}{
\begingroup
\definecolor{FlowBBH}{HTML}{C58B25}
\definecolor{FlowBNS}{HTML}{3678B5}
\definecolor{FlowNSBH}{HTML}{168C87}
\definecolor{FlowInk}{HTML}{203247}
\definecolor{FlowRule}{HTML}{CDD7E1}
\newcommand{\flowNet}[4]{%
 \begin{scope}[shift={(#1,#2)}]
 \foreach \a in {0,1} {\foreach \i in {-1,0,1} {\foreach \j in {-1,0,1} {
  \draw[#3!40,line width=.45pt] (.58*\a,.36*\i)--(.58*\a+.58,.36*\j);
 }}}
 \foreach \a in {0,1,2} {\foreach \i in {-1,0,1} {
  \filldraw[fill=white,draw=#3,line width=.75pt] (.58*\a,.36*\i) circle (.075);
 }}
 \end{scope}}
\begin{tikzpicture}[x=1cm,y=1cm,text=FlowInk,font=\sffamily\fontsize{10}{12}\selectfont,
 line cap=round,line join=round,
 frame/.style={draw=FlowRule,line width=.65pt,rounded corners=1.2mm},
 box/.style={draw=FlowRule,line width=.75pt,rounded corners=1mm,fill=white,align=center,inner sep=4pt},
 title/.style={font=\sffamily\bfseries\fontsize{11}{13}\selectfont,align=center},
 small/.style={font=\sffamily\fontsize{8.5}{10}\selectfont,align=center},
 arrow/.style={-{Stealth[length=1.8mm,width=1.2mm]},line width=.8pt,draw=FlowInk!75},
 sum/.style={circle,draw=FlowBNS,fill=white,minimum size=.48cm,inner sep=0pt,font=\fontsize{13}{14}\selectfont}]
\path[use as bounding box] (0,-.65) rectangle (22,11.55);
\node[title,anchor=west] at (.05,11.25) {b\quad Shared learning and binary-specific waveform generation};
\node[title] at (.95,10.55) {Parameters};
\node[title] at (6.15,10.55) {Shared neural network};
\node[title] at (12.0,10.55) {Specialized networks};
\node[title] at (15.9,10.55) {Waveform\\reconstruction};
\node[title] at (20,10.55) {Generated waveforms};
\draw[frame] (2.4,1.8) rectangle (10.25,10.05);
\node[box,text width=1.42cm,minimum height=2.8cm,font=\fontsize{13}{16}\selectfont] (input) at (.95,5.8)
 {$q$\\[7pt]$\boldsymbol\chi_1,\boldsymbol\chi_2$\\[7pt]$\Lambda_1,\Lambda_2$\\[7pt]$(n_1,n_2)$};
\draw[frame] (2.85,4.45) rectangle (5.65,8.2);
\node[title,font=\sffamily\bfseries\fontsize{10}{12}\selectfont] at (4.25,7.8) {Shared layers};
\foreach \a in {0,1} {\foreach \i in {0,1,2,3} {\foreach \j in {0,1,2,3} {
 \draw[FlowInk!23,line width=.45pt] (3.2+.9*\a,5.0+.65*\i)--(4.1+.9*\a,5.0+.65*\j);
 }}}
\foreach \a in {0,1,2} {\foreach \i in {0,1,2,3} {
 \filldraw[fill=white,draw=FlowInk!75,line width=.8pt] (3.2+.9*\a,5.0+.65*\i) circle (.09);
 }}
\node[small] at (4.25,4.7) {6 residual blocks};
\coordinate (z) at (5.95,5.8);
\draw[arrow] (5.65,5.8)--(z);
\foreach \yy/\col/\name in {9/FlowBBH/BBH,5.8/FlowBNS/BNS,2.8/FlowNSBH/NSBH}{
 \draw[frame,draw=\col!65] (6.65,\yy-.74) rectangle (8.88,\yy+.88);
 \node[title,text=\col,font=\sffamily\bfseries\fontsize{9}{11}\selectfont] at (7.77,\yy+.61) {\name\ branch};
 \flowNet{7.18}{\yy-.05}{\col}{\name}
 \draw[arrow,draw=\col] (z)--(6.15,5.8)--(6.15,\yy)--(6.65,\yy);
 \draw[arrow,draw=\col] (8.88,\yy)--(10.25,\yy);
}

\draw[frame,draw=FlowBNS,line width=1pt] (6.65,5.06) rectangle (8.88,6.68);
\foreach \yy/\col/\name in {9/FlowBBH/BBH,2.8/FlowNSBH/NSBH}{
 \draw[frame,draw=\col,line width=1pt] (10.85,\yy-.88) rectangle (13.18,\yy+.88);
 \node[small,text=\col,font=\sffamily\bfseries\fontsize{9}{11}\selectfont] at (12.02,\yy+.62) {Refinement};
 \flowNet{11.44}{\yy-.03}{\col}{ref\name}
 \draw[arrow,draw=\col] (10.25,\yy)--(10.85,\yy);
 \node[box,draw=\col!60,text width=2.6cm,minimum height=.95cm] (decode-\name) at (15.68,\yy)
 {$\widehat{\mathcal H}=B\,\widehat R$};
 \draw[arrow,draw=\col] (13.18,\yy)--(decode-\name.west);
 \draw[arrow,draw=\col] (decode-\name.east)--(18.28,\yy);
}
\node[box,draw=FlowBNS!60,text width=2.6cm,minimum height=1.05cm] (pre) at (15.68,7.6)
 {Premerger\\[4pt]$\widehat{\mathcal H}=\widehat A_{\rm pk}B\,\widehat R$};
\draw[arrow,draw=FlowBNS] (10.25,5.8)--(10.5,5.8)|-(pre.west);
\draw[frame,draw=FlowBNS,line width=1pt] (10.85,4.35) rectangle (13.18,6.35);
\node[small,text=FlowBNS,font=\sffamily\bfseries\fontsize{9}{10.5}\selectfont] at (12.02,5.98) {Postmerger\\network};
\flowNet{11.44}{5.07}{FlowBNS}{pm}
\draw[arrow,draw=FlowBNS] (12.02,6.35)--(12.02,6.55)--node[above,small] {$\widehat A_{\rm pk}$} (15.68,6.55)--(pre.south);
\fill[FlowInk] (6.15,4.65) circle (.045);
\draw[arrow,draw=FlowBNS] (6.15,4.65)--(10.85,4.65);
\node[small,text=FlowBNS,anchor=south] at (8.55,4.67) {Shared features};
\node[box,draw=FlowInk!60,text width=2.6cm,minimum height=.55cm,font=\sffamily\fontsize{9}{11}\selectfont] (physical) at (3.9,3.85)
 {Fourier features};
\draw[arrow] (input.south)|-(physical.west);
\draw[arrow] (physical.north)--(3.9,4.45);
\draw[arrow,preaction={draw=white,line width=3pt}]
 (physical.east)--(5.95,3.85)
 .. controls (6.04,4.08) and (6.26,4.08) .. (6.35,3.85)
 --(12.02,3.85)--(12.02,4.35);
\draw[arrow,preaction={draw=white,line width=3pt}]
 (9.65,3.85)--(9.65,4.45)
 .. controls (9.88,4.54) and (9.88,4.76) .. (9.65,4.85)
 --(9.65,5.6)
 .. controls (9.88,5.69) and (9.88,5.91) .. (9.65,6.0)
 --(9.65,8.8)
 .. controls (9.88,8.89) and (9.88,9.11) .. (9.65,9.2)
 --(9.65,9.5)--(10.85,9.5);
\draw[arrow] (9.65,3.85)--(9.65,3.3)--(10.85,3.3);
\fill[FlowInk!75] (9.65,3.85) circle (.045);
\node[box,draw=FlowBNS!60,text width=2.6cm,minimum height=1.12cm] (pm) at (15.68,5.23)
 {NRPM + FFT\\[4pt]$\widehat{\mathcal H}$};
\draw[arrow,draw=FlowBNS] (13.18,5.23)--(pm.west);
\node[sum] (bnsadd) at (17.75,5.8) {$+$};
\draw[arrow,draw=FlowBNS] (pre.east)-|(bnsadd.north);
\draw[arrow,draw=FlowBNS] (pm.east)-|(bnsadd.south);
\draw[arrow,draw=FlowBNS] (bnsadd.east)--(18.28,5.8);
\foreach \yy/\col/\name/\lo/\hi in {9/FlowBBH/BBH/-55/22,5.8/FlowBNS/BNS/-3/8,2.8/FlowNSBH/NSBH/-6/3}{
 \node[text=\col,font=\sffamily\bfseries\fontsize{11}{13}\selectfont] at (20,\yy+.97) {\name};
 \begin{scope}[shift={(18.45,\yy)},x=3.15cm,y=.63cm]
  \draw[FlowRule,line width=.4pt] (0,0)--(1,0);
  \draw[\col,line width=.8pt] plot file {figures/generated_overview/\name_generated.dat};
 \end{scope}
 \draw[FlowRule,line width=.45pt] (18.45,\yy-.82)--(21.6,\yy-.82);
 \node[small,anchor=north] at (18.45,\yy-.88) {$\lo$};
 \node[small,anchor=north] at (21.6,\yy-.88) {$\hi$};
 \node[small,anchor=north] at (20.02,\yy-.88) {$t-t_{\rm peak}$ (ms)};
}
\node[small] at (11,1.22) {Thick borders: updated in specialized training};
\draw[FlowRule,line width=.55pt] (.15,.76)--(21.85,.76);
\node[anchor=west,font=\sffamily\fontsize{9.5}{12}\selectfont] at (.2,.32)
 {\textbf{Inputs}\quad $q$: mass ratio\qquad $\boldsymbol\chi_i$: spin\qquad $\Lambda_i$: tidal deformability\qquad $(n_1,n_2)$: binary type};
\node[anchor=west,font=\sffamily\fontsize{9.5}{12}\selectfont] at (.2,-.22)
 {\textbf{Reconstruction}\quad $B$: inspiral baseline\qquad $\widehat R$: spectral residual\qquad $\widehat{\mathcal H}$: generated spectrum};
\end{tikzpicture}
\endgroup}
\caption{\textbf{Physical features and unified waveform generation framework.} \textbf{a,} Reference waveform excerpts for binary black holes (BBHs), binary neutron stars (BNSs), and neutron star black hole (NSBH) systems. \textbf{b,} A shared neural network combines with dedicated networks to generate waveforms. Fourier features feed the shared layers, the BBH and NSBH spectral refinement networks, and the BNS postmerger network. The shared output and the gated output of the selected binary branch are added to obtain the initial prediction; this combination is omitted from the overview for clarity and shown in Fig.~\ref{fig:supp_architecture}. BBH and NSBH refine the spectral coefficients. The BNS waveforms before and after merger share the predicted peak amplitude $\widehat A_{\rm pk}$. The numerical relativity postmerger (NRPM) model generates the postmerger waveform in the time domain, and a fast Fourier transform (FFT) converts it to a spectrum that is added to the premerger spectrum. Symbols are defined below the diagram. After restoring physical units, an inverse Fourier transform gives the waveform in the time domain. The curves on the right are generated waveforms for the same samples as in Fig.~\ref{fig:waveforms}. Each displayed excerpt is normalized by its own peak, and time is measured relative to the amplitude peak of the corresponding reference waveform. Nodes and connections illustrate the module structure. Gold, blue, and teal denote BBH, BNS, and NSBH, respectively. The detailed architecture and training settings are given in Fig.~\ref{fig:supp_architecture} and Materials and Methods.}
\label{fig:architecture}
\end{figure*}
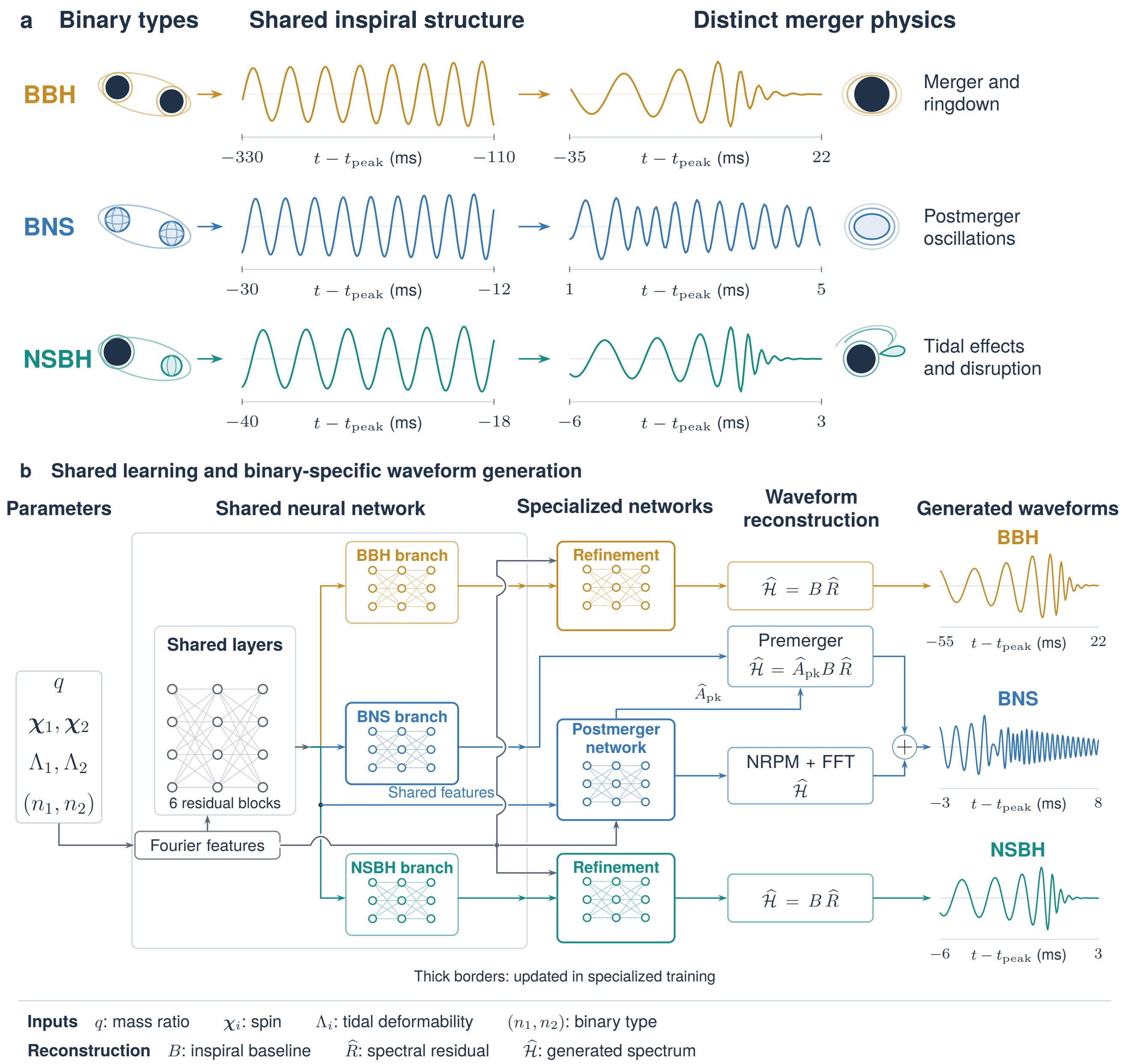

URSA-GW uses inspiral laws shared by the three compact binary types to construct a waveform representation and describes their waveform features through shared learning and branches specific to each binary type (Fig.~\ref{fig:architecture}). We divide each reference spectrum by a Newtonian analytic baseline to remove the leading amplitude and phase variations during inspiral, then compress the logarithmic amplitude and unwrapped phase of the residual spectrum into low dimensional coefficients. The network predicts these coefficients to recover the full spectrum. The shared layers are trained jointly on waveforms from the three binary types to exploit common relations between parameters and waveforms, while dedicated branches describe differences between binary types. For BBH and NSBH, spectral refinement networks further improve coefficient predictions. Because BNS postmerger radiation arises from remnant oscillations, we model it separately from the waveform before merger. The network predicts postmerger physical parameters that drive the numerical relativity postmerger (NRPM) model to generate the radiation, whose spectrum is then added to the reconstructed premerger spectrum to obtain the complete BNS waveform.

The network inputs comprise the mass ratio $q$, the dimensionless spin vectors $\boldsymbol\chi_i$ of the two objects, their dimensionless tidal deformabilities $\Lambda_i$, and the binary type code $(n_1,n_2)$, where $i=1,2$. We train and evaluate the model using polarization waveforms for a fixed viewing direction at a fixed reference total mass for each binary type. The ranges of $q$, $\boldsymbol\chi_i$, and $\Lambda_i$ are listed in Table~\ref{tab:domain}. For vacuum BBHs, waveforms at different total masses can be obtained by mass scaling, covering stellar origin binary black hole (SOBBH), intermediate mass binary black hole (IMBBH), and supermassive binary black hole (SMBBH) systems. The detailed network architecture is shown in Fig.~\ref{fig:supp_architecture}. Reference waveforms, spectral representations, training, and decoding are described in Materials and Methods.

\begin{table*}[tbp]
\centering\small
\caption{\textbf{Physical parameter domains, radiation content and waveform coverage.} Parameter domains, reference waveform models and their physical effects, physical frequency ranges, waveform cycles, and durations for the three compact binary types.}
\label{tab:domain}
\begingroup
\setlength{\tabcolsep}{2.8pt}
\renewcommand{\arraystretch}{1.25}
\renewcommand{\tabularxcolumn}[1]{m{#1}}
\newcommand{\MassSlice}[1]{\parbox[c][1.10cm][c]{\linewidth}{\centering #1}}
\begin{tabularx}{\textwidth}{@{}>{\centering\arraybackslash}m{1.15cm}>{\centering\arraybackslash}m{2.8cm}>{\centering\arraybackslash}X>{\centering\arraybackslash}m{1.45cm}>{\centering\arraybackslash}m{2.7cm}>{\centering\arraybackslash}m{1.4cm}>{\centering\arraybackslash}m{2.15cm}@{}}
\toprule
\shortstack{Binary\\type} & Intrinsic domain & Reference waveform and radiation & \shortstack{Mass class\\$M/M_\odot$} & \shortstack{Frequency band\\$[f_{\min},f_{\max}]$ (Hz)} & \shortstack{Waveform\\cycles} & Duration range\\
\midrule
BBH
 & $q\in[1,10]$\newline $|\bm\chi_{1,2}|\in[0,0.9]$\newline $(\Lambda_1,\Lambda_2)=(0,0)$
 & SEOBNRv5PHM\newline Precession + higher modes\newline $(\ell,m)=(2,2),(2,1),(3,3),$\newline $(3,2),(4,4),(4,3),(5,5)$
 & \MassSlice{SOBBH\\$10$--$10^2$}\par\MassSlice{IMBBH\\$10^2$--$10^5$}\par\MassSlice{SMBBH\\$10^5$--$10^9$}
 & \MassSlice{$[2.03,$\\$4.06\times10^{3}]$}\par\MassSlice{$[2.03\times10^{-3},$\\$4.06\times10^{2}]$}\par\MassSlice{$[2.03\times10^{-7},$\\$4.06\times10^{-1}]$}
 & 596--2249
 & \MassSlice{$18.4\,\mathrm{s}$--$10.0\,\mathrm{min}$}\par\MassSlice{$3.06\,\mathrm{min}$--$6.96\,\mathrm{d}$}\par\MassSlice{$2.13\,\mathrm{d}$--$190.5\,\mathrm{yr}$}\\
\midrule
BNS & $q\in[1,1.5]$\newline $|\bm\chi_{1,2}|\in[0,0.5]$\newline $\Lambda_{1,2}\in[0,1000]$
 & IMRPhenomXP\_\allowbreak NRTidalv3 + NRPM\newline Precession + tides\newline Postmerger radiation
 & $2.8$ & $[72.51,14501.82]$ & 652--5804 & $5.72$--$6.13\,\mathrm{s}$\\
\addlinespace[9pt]
NSBH & $q\in[2.5,59/11]$\newline $|\bm\chi_1|\in[0,0.9]$\newline $|\bm\chi_2|\in[0,0.1]$\newline $\Lambda_1=0$\newline $\Lambda_2\in[0,1000]$
 & TEOBResumS-Giotto-BHNS\newline Precession + higher modes\newline Tides and disruption\newline $(\ell,m)=(2,2),(3,3),(4,4)$
 & $7$ & $[29.00,5800.73]$ & 717--3373 & $15.8$--$25.8\,\mathrm{s}$\\[3pt]
\bottomrule
\end{tabularx}
\par\vspace{5pt}
\begin{minipage}{\textwidth}
\footnotesize
The frequency ranges are obtained by converting the dimensionless output grid $u=GMf/c^3\in[10^{-3},0.2]$ to physical frequencies, where $f$ is the physical frequency and $M$ is the binary total mass. The reference total masses for BBH, BNS, and NSBH are $60M_\odot$, $2.8M_\odot$, and $7M_\odot$, respectively. Each BBH mass class lists the lowest and highest frequencies across its mass interval. At a single mass, the frequency band is $[3.38,676.75]\times(60M_\odot/M)\,\mathrm{Hz}$. Duration scales linearly with $M$, while the number of waveform cycles remains unchanged~\cite{Purrer2014}. At $60M_\odot$, the duration is 110.19--360.70\,s. The mass classes are stellar origin binary black hole (SOBBH), intermediate mass binary black hole (IMBBH), and supermassive binary black hole (SMBBH) systems. Cycle and duration ranges are calculated from 100,000 reference waveforms per binary type. BNS records include 0.5\,s after merger.
\end{minipage}
\endgroup
\end{table*}

\begin{figure}[tbp]
\centering
\includegraphics[width=\columnwidth]{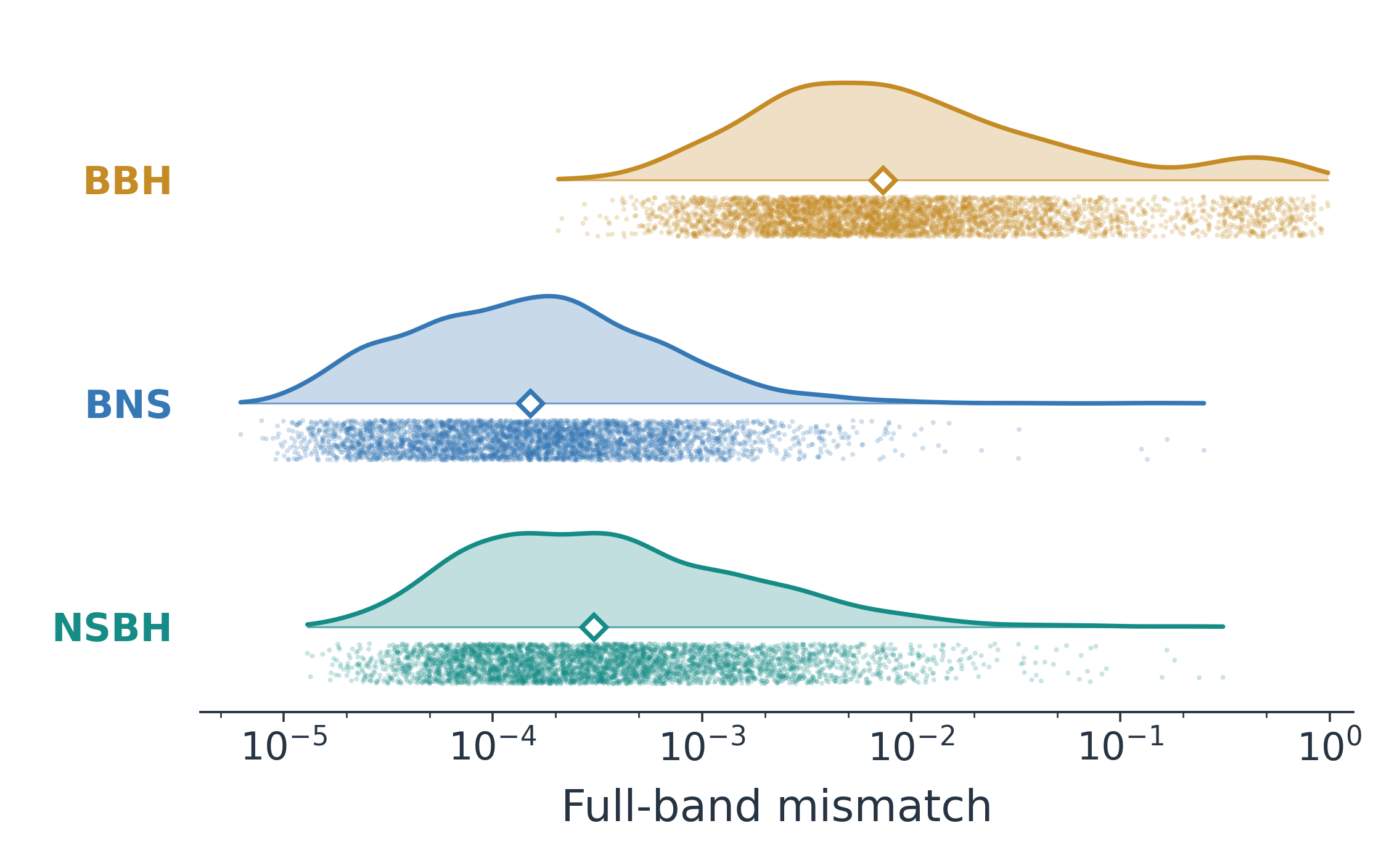}
\caption{\textbf{Waveform mismatch across three binary types.} Distributions of full band mismatch for 4096 test waveforms per binary type, BBH, BNS, and NSBH. The mismatch is computed without detector noise weighting. Curves show probability densities, points represent individual test waveforms, and white diamonds mark medians. Gold, blue, and teal denote BBH, BNS, and NSBH, respectively. Summary statistics are given in Table~\ref{tab:mismatch}.}
\label{fig:mismatch}
\label{fig:performance}
\end{figure}

\subsection{Accuracy of generated waveforms}
We evaluate the generation accuracy of URSA-GW on 4096 test waveforms for each of BBH, BNS, and NSBH. Waveform mismatch measures the difference between generated and reference waveforms using frequency integration weights without detector noise weighting, as defined in Materials and Methods. The mean mismatches are $5.34\times10^{-2}$, $5.89\times10^{-4}$, and $1.63\times10^{-3}$ for the three binary types, respectively. Their distributions and summary statistics are shown in Fig.~\ref{fig:mismatch} and Table~\ref{tab:mismatch}. The BBH mismatch distribution has a long tail toward high values. Samples with high mismatch are concentrated mainly at high mass ratios, raising the mean above the median and identifying this region as a focus for further improvements in BBH generation accuracy.

\begin{table}[tbp]
\centering\small
\caption{\textbf{Full band waveform mismatch.} Mean and median full band mismatches relative to the reference waveforms for 4096 test waveforms per binary type, BBH, BNS, and NSBH.}
\label{tab:mismatch}
\begin{tabular*}{\columnwidth}{@{\extracolsep{\fill}}l S[table-format=1.2e-1] S[table-format=1.2e-1]@{}}
\toprule
Binary type & {Mean} & {Median}\\
\midrule
BBH & 5.34e-2 & 7.35e-3\\
BNS & 5.89e-4 & 1.51e-4\\
NSBH & 1.63e-3 & 3.04e-4\\
\bottomrule
\end{tabular*}
\end{table}

\begin{figure*}[tbp]
\centering
\includegraphics[width=0.90\textwidth]{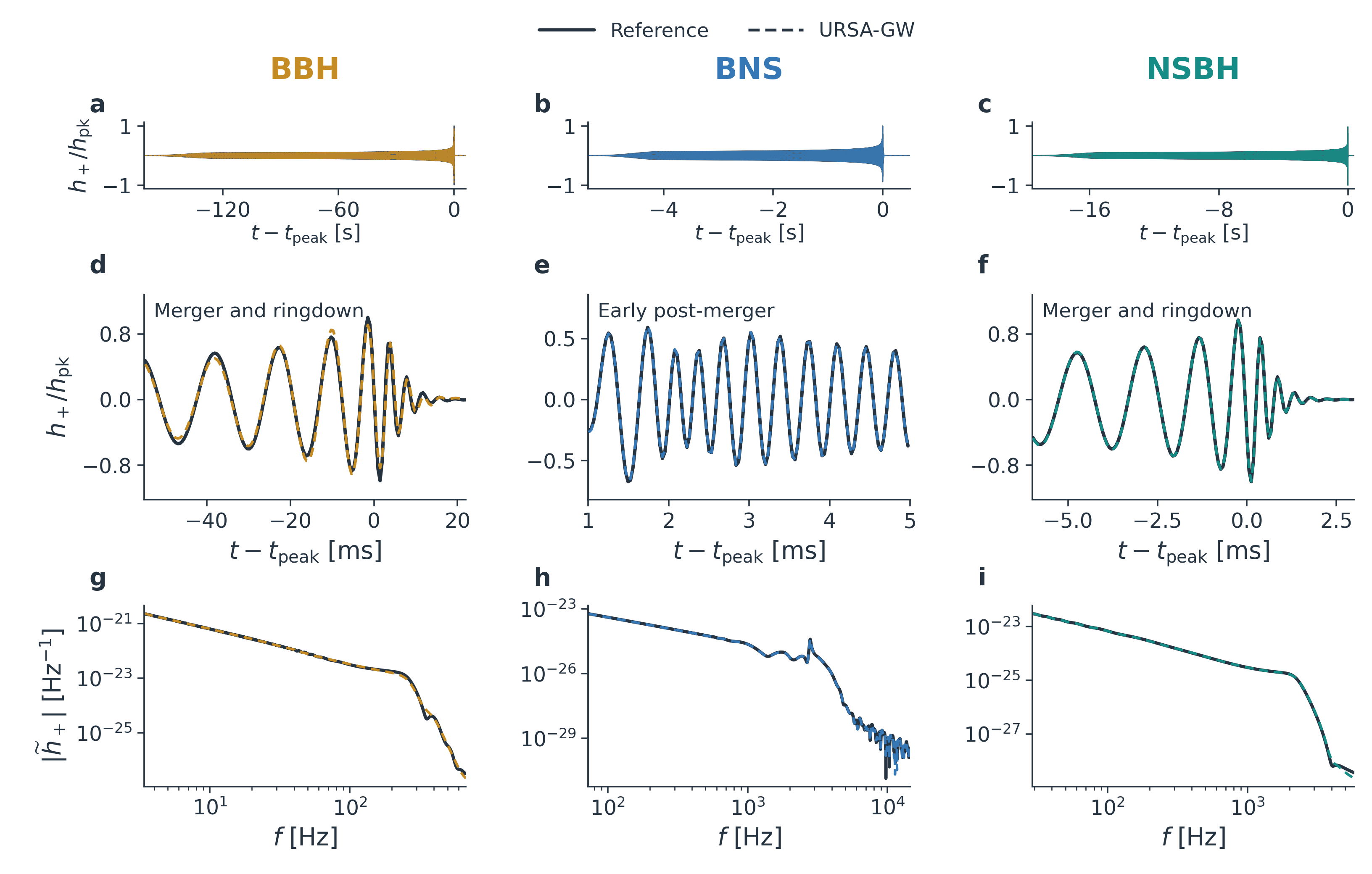}
\caption{\textbf{Waveform recovery in the time and frequency domains.} Columns correspond to BBH, BNS, and NSBH. Solid lines show reference waveforms, and colored dashed lines show URSA-GW waveforms. \textbf{a--c,} Complete waveforms in the time domain. \textbf{d--f,} Waveforms near merger for BBH and NSBH and early postmerger oscillations for BNS. \textbf{g--i,} Comparison of spectral amplitudes. All waveforms in the time domain are normalized by the peak absolute strain $h_{\rm pk}$ of the reference waveform. Sample parameters, waveform cycles, durations, and mismatches are listed in Table~\ref{tab:examples}.}
\label{fig:waveforms}
\label{fig:spectra}
\end{figure*}

\subsection{Waveform recovery in the time and frequency domains}
Figure~\ref{fig:waveforms} shows three waveforms selected from the BBH, BNS, and NSBH test sets. Their parameters and mismatches are listed in Table~\ref{tab:examples}, and the selection procedure is described in Materials and Methods. The generated and reference waveforms show consistent evolution on a common time axis, covering BBH inspiral, merger, and ringdown, NSBH inspiral and merger with tidal and disruption effects, and BNS inspiral and postmerger radiation. Enlarged views further show recovery of oscillations near merger for BBH and NSBH and of the amplitude and phase of early BNS postmerger oscillations (Fig.~\ref{fig:waveforms}d--f). The BNS waveform includes a 0.5\,s postmerger record generated by NRPM using physical parameters predicted by the network. The frequency domain comparison for the same samples shows that generated waveforms also reproduce the amplitude variations of the reference waveforms across the adopted band (Fig.~\ref{fig:waveforms}g--i). Together, the time and frequency domain results illustrate waveform recovery for different binary types.

\begin{table*}[tbp]
\centering\small
\caption{\textbf{Parameters and properties of the waveforms shown in Fig.~\ref{fig:waveforms}.} Physical parameters, waveform cycles, durations, and full band mismatches of the three test samples. All distances are 100\,Mpc, and $\boldsymbol\chi_i$ is given in Cartesian components. The BNS duration includes 0.5\,s of postmerger radiation.}
\label{tab:examples}
\setlength{\tabcolsep}{2.8pt}
\begin{tabularx}{\textwidth}{@{}l S[table-format=2.2] S[table-format=2.2] >{\centering\arraybackslash}X >{\centering\arraybackslash}X c S[table-format=3.2] S[table-format=3.2] S[table-format=1.2e-1]@{}}
\toprule
\shortstack{Binary\\type}&{$m_1/M_\odot$}&{$m_2/M_\odot$}&$\boldsymbol\chi_1$&$\boldsymbol\chi_2$&$(\Lambda_1,\Lambda_2)$&{\shortstack{Waveform\\cycles}}&{\shortstack{Duration\\/ s}}&{mismatch}\\
\midrule
BBH&46.60&13.40&$(-0.073,-0.057,-0.021)$&$(-0.003,-0.076,0.130)$&(0.0,0.0)&936.96&167.10&3.26e-4\\
BNS&1.48&1.32&$(0.126,0.074,0.267)$&$(0.019,0.016,-0.012)$&(520.0,798.0)&880.89&5.88&1.19e-5\\
NSBH&5.53&1.47&$(-0.192,0.174,-0.065)$&$(0.003,-0.044,0.000)$&(0.0,565.6)&904.88&19.94&2.27e-5\\
\bottomrule
\end{tabularx}
\end{table*}

Statistics from 100,000 training reference waveforms per binary type show that the waveform records span hundreds to thousands of cycles within the adopted band (Table~\ref{tab:domain}). At reference total masses of $60M_\odot$, $2.8M_\odot$, and $7M_\odot$ for BBH, BNS, and NSBH, the output frequency ranges are 3.38--676.75\,Hz, 72.51--14501.82\,Hz, and 29.00--5800.73\,Hz, and the corresponding band limited waveform durations are 110.19--360.70\,s, 5.72--6.13\,s, and 15.8--25.8\,s, respectively. The BBH waveforms cover approximately 1.8--6.0 minutes of evolution, while generating one complex spectrum with 4480 points on the GPU takes approximately 5.2\,ms. The corresponding generation times for BNS and NSBH are approximately 5.0\,ms and 4.1\,ms (Table~\ref{tab:speed}).

For vacuum BBHs, scaling the $60\,M_\odot$ reference waveforms to $10$--$10^9\,M_\odot$ while holding $q$, $\boldsymbol\chi_i$, and the dimensionless frequency band fixed gives band limited durations from approximately 18.4\,s to 190.5 years, with the number of waveform cycles unchanged. This mass scaling allows the same waveform representation to cover stellar mass, intermediate mass, and supermassive BBHs. The duration definition and scaling relation are given in Materials and Methods.

\begin{figure}[tbp]
\centering
\includegraphics[width=\columnwidth]{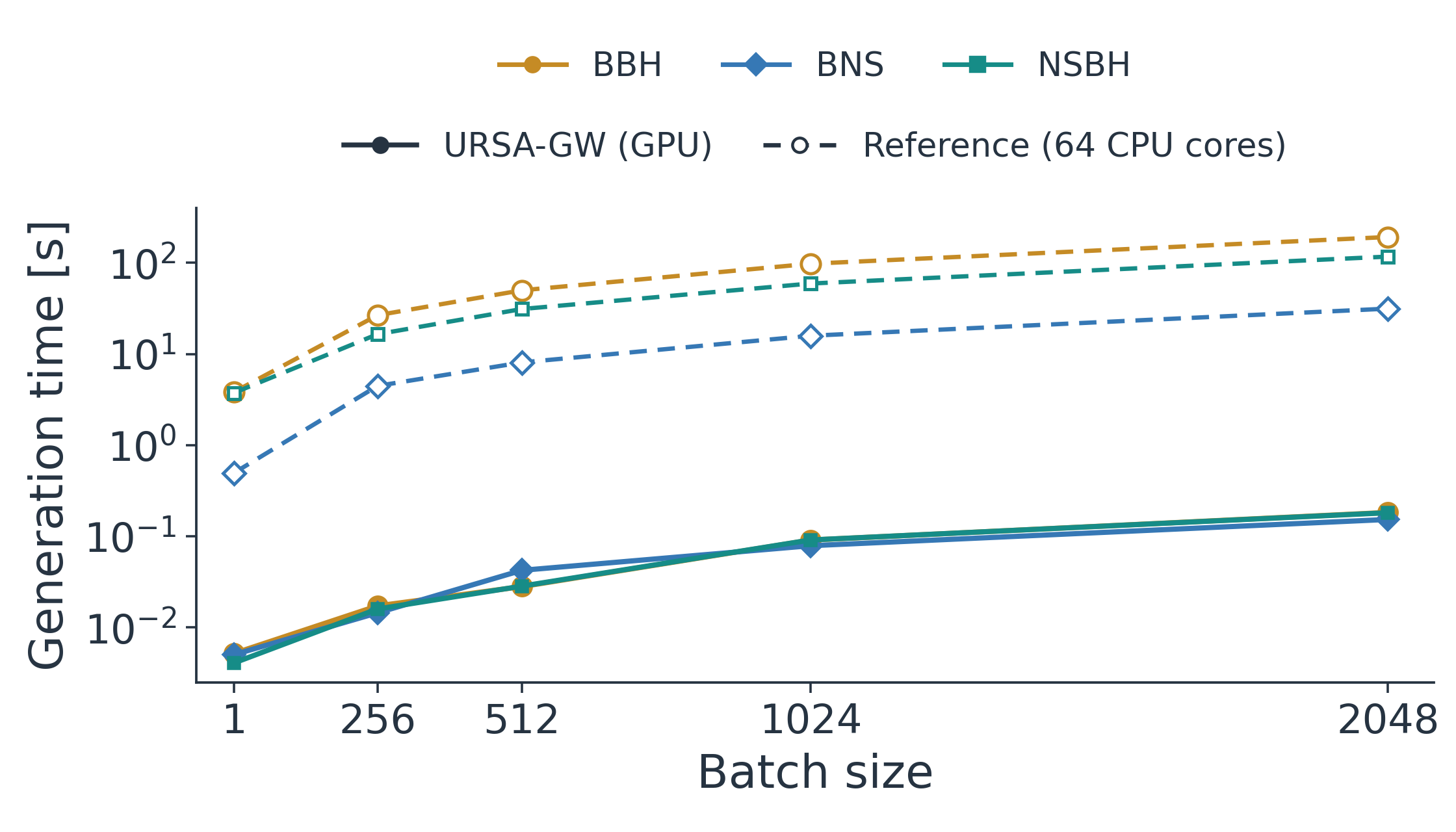}
\caption{\textbf{Computational efficiency of waveform generation.} Time to generate an entire batch at different batch sizes. Solid lines and filled points show URSA-GW calculations on the GPU, while dashed lines and open points show reference calculations on 64 CPU cores. Gold, blue, and teal denote BBH, BNS, and NSBH, respectively. The vertical axis is logarithmic. Timings and speedup factors are given in Table~\ref{tab:speed}.}
\label{fig:speed}
\end{figure}

\subsection{Computational efficiency of waveform generation}
URSA-GW substantially reduces the computation time for both individual calls and batch generation across all three binary types (Fig.~\ref{fig:speed} and Table~\ref{tab:speed}). Generation uses cache reuse and batch decoding on the GPU, as described in the Computational optimization and efficiency tests subsection of Materials and Methods. Relative to reference calculations on 64 CPU cores, timing covers the complete procedure from intrinsic parameters to the final waveform output. At a batch size of 512, generation takes approximately 0.028\,s for both BBH and NSBH, corresponding to speedup factors of approximately 1784 and 1099, respectively. BNS waveform generation includes postmerger radiation and takes approximately 0.042\,s for a batch of 512, giving a speedup of approximately 191.

\begin{table*}[tbp]
\centering\small
\caption{\textbf{Waveform generation time and speedup across batch sizes.} CPU and GPU columns give the time in seconds to generate an entire batch using the reference models on 64 physical CPU cores and URSA-GW on an NVIDIA A800 GPU, respectively. The speedup is the reference model time divided by the URSA-GW time.}
\label{tab:speed}
\begingroup
\setlength{\tabcolsep}{3pt}
\renewcommand{\arraystretch}{1.3}
\begin{tabular*}{\textwidth}{@{\extracolsep{\fill}}r *{3}{S[table-format=3.3] S[table-format=1.4] S[table-format=4.2]}@{}}
\toprule
& \multicolumn{3}{c}{\textbf{BBH}} & \multicolumn{3}{c}{\textbf{BNS}} & \multicolumn{3}{c}{\textbf{NSBH}}\\
\cmidrule(lr){2-4}\cmidrule(lr){5-7}\cmidrule(l){8-10}
Batch size & {CPU / s} & {GPU / s} & {Speedup} & {CPU / s} & {GPU / s} & {Speedup} & {CPU / s} & {GPU / s} & {Speedup}\\
\midrule
1 & 3.819 & 0.0052 & 738.97 & 0.494 & 0.0050 & 98.54 & 3.738 & 0.0041 & 919.97\\
256 & 26.650 & 0.0172 & 1547.67 & 4.466 & 0.0144 & 309.28 & 16.498 & 0.0159 & 1036.20\\
512 & 50.012 & 0.0280 & 1784.31 & 8.083 & 0.0423 & 191.09 & 31.159 & 0.0283 & 1099.13\\
1024 & 97.832 & 0.0908 & 1077.10 & 15.860 & 0.0788 & 201.37 & 59.326 & 0.0911 & 651.56\\
2048 & 192.211 & 0.1825 & 1053.02 & 31.462 & 0.1529 & 205.77 & 117.180 & 0.1812 & 646.62\\
\bottomrule
\end{tabular*}
\endgroup
\end{table*}

Generating one waveform on the GPU takes approximately 4.1--5.2\,ms, with speedup factors of approximately 739, 99, and 920 for BBH, BNS, and NSBH, respectively. At a batch size of 2048, generation takes less than 0.2\,s for all three binary types, compared with 31.46--192.21\,s for the corresponding reference calculations. Results for individual batch sizes are listed in Table~\ref{tab:speed}. Rapid individual waveform evaluations can reduce repeated computational costs in parameter inference, while batch generation can support large scale waveform simulations and the construction of simulated datasets. These results suggest that URSA-GW may provide an efficient waveform generation method for large scale event searches, rapid parameter inference, and population studies with next generation gravitational wave observations.

\section{Discussion}
This work shows that BBH, BNS, and NSBH waveform generation can be implemented within a single framework while retaining spin precession, higher order radiation modes, and neutron star matter effects present in the reference models of the respective binary types. Existing neural network surrogate models have advanced rapid generation of BBH and BNS waveforms~\cite{Chua2019ROM,ANNSur2021,Liao2021CAE,mlgwBNS2023,Grimbergen2024HM,Sun2025cAE,Shi2025Eccentric,Shi2025HigherModes,Freitas2025DANSur,Whittall2026}. Building on these developments, our framework incorporates the inspiral evolution shared by the three binary types and their distinct merger and postmerger radiation into a unified waveform representation and generation procedure. The physical basis for this unification is that inspiral in all three systems is driven by gravitational radiation, whereas tidal effects, neutron star disruption, and remnant oscillations require their own physical descriptions. URSA-GW represents the common inspiral trend with a shared analytic baseline, then retains these differences through spectral representations and physical decoding for the respective binary types, allowing one model to describe their waveform features while maintaining computational efficiency.

For binaries containing neutron stars, matter effects in the waveform provide important probes of neutron star internal structure and the equation of state. In NSBH systems, $q$, $\boldsymbol\chi_1$, and neutron star compactness jointly affect tidal disruption and modify the radiation at high frequencies near merger~\cite{Foucart2018Remnant,TEOBBHNS}. Oscillations of BNS merger remnants reflect the properties of dense matter. Phase transitions may alter characteristic postmerger frequencies~\cite{Bauswein2019PhaseTransition}, while finite temperature effects influence remnant structure and evolution~\cite{Oechslin2007EOS}. URSA-GW retains the tidal, disruption, and postmerger effects of the selected reference models, allowing rapid waveform generation to describe both the tidal response during inspiral and subsequent remnant oscillations. With reference models containing a broader range of matter effects, the framework could be used to simulate waveforms for different equations of state, providing efficient waveform calculations for studying neutron star matter and its observational signatures.

For complex amplitude and phase variations caused by precession and higher order modes, improvements in the spectral representation and the mapping from parameters to coefficients may further improve the description of these structures. Cache reuse and batch calculations on the GPU reduce the cost of the current decoding procedure and improve the overall efficiency of generation from binary parameters to waveform output.

The modular design of URSA-GW provides a basis for extension to the frequency bands of space based gravitational wave observations. For earlier inspiral waveforms at lower frequencies, the overall framework of an analytic baseline and shared learning can be retained, while the frequency sampling and residual spectral representation are adjusted. New or modified dedicated modules can then be trained on corresponding reference waveforms to incorporate the extended band into the unified generation procedure. This architecture also allows input features and decoding modules to be adapted to new physical processes, providing a way to extend frequency coverage and the description of physical effects.

The current evaluation concerns polarization waveforms for a fixed viewing direction within the specified parameter domain of each binary type. BBH total mass coverage follows from mass scaling of vacuum waveforms, while BNS and NSBH are modeled at fixed reference total masses. The mismatch used here measures agreement with the selected reference models. The long tail toward high BBH mismatches indicates that coefficient predictions still require improvement in some parameter regions. The framework can also be extended to other viewing directions and broader mass ranges.

URSA-GW can be used for large scale waveform simulations and the construction of simulated datasets, and provides a rapid generation tool for parameter inference requiring repeated waveform evaluations. It could be combined with Bayesian sampling~\cite{Ashton2019,RomeroShaw2020,Smith2020Parallel} and neural posterior estimation~\cite{Gabbard2022,Dingo2021} to reduce the cost of waveform evaluations in likelihood calculations and importance reweighting~\cite{Dax2023}. Beyond the three binary types studied here, the framework could be extended to eccentric binaries and extreme mass ratio inspirals (EMRIs), with appropriate waveform representations describing the amplitude and phase modulations caused by eccentric orbits and the multiple harmonics in long duration EMRI waveforms.

\section{Materials and Methods}
\label{sec:methods}

\subsection{Reference waveforms and dataset construction}
The reference dataset contains three binary types, BBH, BNS, and NSBH, with reference total masses fixed at $60\,M_\odot$, $2.8\,M_\odot$, and $7\,M_\odot$, respectively. We define $q=m_1/m_2\geq1$. Given the total mass $M$ and $q$, the component masses are $m_1=Mq/(1+q)$ and $m_2=M/(1+q)$. For BBH, $q\in[1,10]$ covers systems from equal masses to strong mass asymmetry, allowing the model to be evaluated across waveform structures at different $q$. For BNS, $q\in[1,1.5]$ matches the $q$ calibration range of the adopted NRPM postmerger model~\cite{NRPM2019}. For NSBH, $q\in[2.5,59/11]$ at a fixed total mass of $7\,M_\odot$ corresponds to neutron star masses of $1.1$--$2.0\,M_\odot$ and black hole masses of $5.0$--$5.9\,M_\odot$. For each binary type, $q$ is sampled uniformly over its prescribed interval. Training, validation, and test parameters follow the same sampling distributions.

The spin magnitudes $|\boldsymbol\chi_i|$ of the two objects are sampled independently and uniformly within their prescribed intervals, with isotropically distributed directions of $\boldsymbol\chi_i$. For BBH, both $|\boldsymbol\chi_1|$ and $|\boldsymbol\chi_2|$ lie in $[0,0.9]$. For BNS, both $|\boldsymbol\chi_1|$ and $|\boldsymbol\chi_2|$ lie in $[0,0.5]$. For NSBH, $|\boldsymbol\chi_1|$ and $|\boldsymbol\chi_2|$ lie in $[0,0.9]$ and $[0,0.1]$, respectively. For a neutron star, $\Lambda_i=(2/3)k_{2,i}C_i^{-5}$, where $k_{2,i}$ is the quadrupolar Love number and $C_i=Gm_i/(R_i c^2)$ is the compactness~\cite{Flanagan2008Tidal,Hinderer2008Love}. Black holes have $\Lambda_i=0$, while neutron star values of $\Lambda_i$ are sampled independently and uniformly in $[0,1000]$. For BNS, $\Lambda_1$ and $\Lambda_2$ need not correspond to the same equation of state and are not ordered, allowing the tidal responses of the two neutron stars to be described independently.

To give all three binary types a common input format, we combine $q$, $\boldsymbol\chi_i$, $\Lambda_i$, and $n_i$ into the 11 dimensional vector in Eq.~\eqref{eq:intrinsic_input},
\begin{equation}
\boldsymbol{x}=\left(q,\boldsymbol\chi_1,\boldsymbol\chi_2,\Lambda_1,\Lambda_2,n_1,n_2\right)\in\mathbb R^{11},
\label{eq:intrinsic_input}
\end{equation}
where $\boldsymbol\chi_i=(\chi_{ix},\chi_{iy},\chi_{iz})$ is the dimensionless spin vector, and $n_i=0$ and 1 denote a black hole and a neutron star, respectively. The binary type codes for BBH, BNS, and NSBH are $(0,0)$, $(1,1)$, and $(0,1)$, respectively. The model selects the corresponding branch using $(n_1,n_2)$. All reference waveforms have a luminosity distance of $100\,\mathrm{Mpc}$, zero inclination, and the same fixed reference phase. These three quantities are not network inputs. To represent different binary types on a common frequency grid, we make the frequency and frequency domain strain dimensionless using the reference total mass of each binary type. Defining $M_t=GM/c^3$ and $d_t=d_L/c$, Eq.~\eqref{eq:dimensionless_waveform} relates physical frequency and the Fourier transform $\widetilde h(f)$ of the plus polarization $h_+(t)$ to the dimensionless frequency $u$ and spectrum $\mathcal H(u)$,
\begin{equation}
u=M_t f,\qquad
\widetilde h(f)=\frac{M_t^2}{d_t}\mathcal H(u).
\label{eq:dimensionless_waveform}
\end{equation}

BBH reference waveforms are generated with SEOBNRv5PHM~\cite{SEOBNRv5PHM}, including spin precession and the radiation modes $(\ell,m)=(2,2)$, $(2,1)$, $(3,3)$, $(3,2)$, $(4,4)$, $(4,3)$, and $(5,5)$. Modes in the coprecessing frame and their precession evolution are used to construct polarization waveforms for a fixed viewing direction. The time domain sampling rate is at least 4096\,Hz and is increased as required by the highest frequency in the target band. NSBH reference waveforms use TEOBResumS-Giotto-BHNS~\cite{TEOBBHNS}, including spin precession, tidal effects, and tidal disruption, with the $(2,2)$, $(3,3)$, and $(4,4)$ radiation modes retained and a time domain sampling rate of 32768\,Hz. The connection between inspiral and ringdown for each mode follows the reference model settings. At the few sample positions before the attachment point of the $(3,3)$ mode where evaluation of the ringdown formula outside its domain gives nonfinite values, the corresponding EOB inspiral values are retained. BNS reference waveforms join a premerger waveform described by IMRPhenomXP\_NRTidalv3 to a postmerger waveform described by NRPM~\cite{IMRPhenomXPHM,NRTidalv3,NRPM2019}. Interpolation near the amplitude peak of the premerger waveform determines the peak time, peak amplitude, phase, instantaneous frequency, and its derivative, which set the postmerger initial conditions. Subsequent amplitude, time, and frequency nodes follow the NRPM fitting relations. The postmerger time window is 0.5\,s, with a sampling rate of 32768\,Hz. The adopted NRPM implementation uses a branch threshold of 60. When the tidal coupling parameter defined in Eq.~\eqref{eq:kappa}
\begin{equation}
\kappa_2^T=\frac{3q(q^3\Lambda_1+\Lambda_2)}{(1+q)^5}
\label{eq:kappa}
\end{equation}
is below this threshold, only the short transition from merger to the first amplitude minimum is retained. This branch is determined by the input parameters and applied during waveform decoding.

The dataset for each binary type contains $10^6$ training samples, 4096 validation samples, and 4096 test samples. Sample parameters and data partitions are fixed before reference waveform generation. The training set is expanded in stages to 8192, $10^5$, and $10^6$ samples per binary type, with each stage retaining all samples from the preceding stage. The spectral compression bases, normalization parameters of the prediction targets, and loss scale parameters are determined from the initial 8192 training samples per binary type and remain fixed during subsequent data expansion and training. The validation set is used for model selection. The test set is used only for the final assessment of waveform agreement and does not participate in training or model selection.

\subsection{Analytic baseline and residual spectral compression}
Waveforms of all three binary types are represented over the dimensionless frequency interval $u\in[10^{-3},0.2]$. The low frequency interval $[10^{-3},0.012)$ contains 384 logarithmically spaced nodes, and the high frequency interval $[0.012,0.2]$ contains 4096 uniformly spaced nodes, giving 4480 points in total. The three binary types share the same dimensionless frequency samples. Their physical frequency bands differ because each type uses a different fixed reference total mass, while all samples within a binary type have identical physical frequency nodes. The leading amplitude and phase evolution during inspiral can be obtained from the stationary phase approximation~\cite{Cutler1994}. To reduce the amplitude and phase variations that must be represented and learned, we first divide the dimensionless frequency domain waveform by the Newtonian analytic baseline
\begin{equation}
\begin{split}
B(u;q)&=A_N(u;\eta)e^{i\Phi_N(u;\eta)},\\
A_N(u;\eta)&=\sqrt{\frac{5}{24}}\pi^{-2/3}\sqrt{\eta}\,u^{-7/6},\\
\Phi_N(u;\eta)&=-\frac{3}{128\eta}(\pi u)^{-5/3},\\
\eta&=\frac{q}{(1+q)^2}.
\end{split}
\label{eq:newtonian_base}
\end{equation}
In Eq.~\eqref{eq:newtonian_base}, $A_N$ and $\Phi_N$ are the Newtonian inspiral amplitude and phase, and $\eta$ is the symmetric mass ratio. The analytic baseline $B$ describes the leading inspiral trend shared by the three binary types. We define the residual spectrum as the ratio of the reference spectrum to the baseline, $R(u)=\mathcal H(u)/B(u;q)$. Taking this ratio removes both the baseline amplitude scale and its leading phase, so the residual describes corrections to the known inspiral trend through an amplitude ratio and a phase difference. During generation, multiplying the predicted residual spectrum by the analytic baseline gives $\widehat{\mathcal H}(u)=B(u;q)\widehat R(u)$. For BNS, the premerger spectrum is also normalized by the merger peak amplitude $A_{\rm pk}$ to separate the overall amplitude scale from the spectral shape. Spectral coefficients are constructed from $R_{\rm pre}(u)=\mathcal H_{\rm pre}(u)/[A_{\rm pk}B(u;q)]$, and the amplitude is restored during generation through $\widehat{\mathcal H}_{\rm pre}(u)=\widehat A_{\rm pk}B(u;q)\widehat R_{\rm pre}(u)$. Here $A_{\rm pk}$ is the amplitude of the dimensionless time domain analytic waveform at the merger peak. Its prediction is supplied by the postmerger parameter network, so the waveforms before and after merger share the same peak amplitude scale. Postmerger radiation is represented separately. Weighted principal component analysis (PCA)~\cite{Tiglio2022} is applied to the logarithmic amplitude and unwrapped phase of the residual spectrum in each frequency interval. We retain 32, 64, 64, and 128 principal components for the low frequency logarithmic amplitude, low frequency phase, high frequency logarithmic amplitude, and high frequency phase, respectively, giving 288 spectral coefficients. Each binary type uses means and bases determined from its own training data, with the same coefficient dimensions and ordering. High frequency weights depend on the mean squared waveform amplitude and the frequency integration spacing, prioritizing spectral structures with larger amplitudes during compression.

The logarithmic amplitude and unwrapped phase of the low and high frequency residuals are encoded and decoded using Eq.~\eqref{eq:pca_decode},
\begin{equation}
\begin{split}
\boldsymbol c_r&=(\boldsymbol w_r\odot\boldsymbol y_r-\boldsymbol\mu_r)V_r^{\mathsf T},\\
\widehat{\boldsymbol y}_r&=(\widehat{\boldsymbol c}_r V_r+\boldsymbol\mu_r)\oslash\boldsymbol w_r,
\end{split}
\label{eq:pca_decode}
\end{equation}
where $r$ denotes one of the four spectral components. During encoding, the residual component $\boldsymbol y_r$ is weighted by $\boldsymbol w_r$, the weighted training mean $\boldsymbol\mu_r$ is subtracted, and the result is projected onto the truncated orthonormal basis $V_r$ to obtain the low dimensional coefficients $\boldsymbol c_r$. During generation, the network predictions $\widehat{\boldsymbol c}_r$ are multiplied by the basis, the mean is restored, and the result is divided by the weights to recover the logarithmic amplitude or unwrapped phase $\widehat{\boldsymbol y}_r$. The symbols $\odot$ and $\oslash$ denote elementwise multiplication and division. The recovered logarithmic amplitude and phase are combined into a complex residual spectrum and multiplied by the analytic baseline in Eq.~\eqref{eq:newtonian_base}. For the BNS premerger part, the predicted peak amplitude $\widehat A_{\rm pk}$ is also restored to obtain the generated spectrum. BBH and NSBH use a common phase convention before encoding. For BNS, the amplitude peak is placed at $t=0$, and the phase of the analytic waveform at that time is set to zero.

\subsection{Physical decoding of BNS postmerger radiation}
BNS postmerger (PM) radiation is represented by 16 quantities comprising 15 continuous physical parameters and one branch flag. The continuous parameters include the peak amplitude, four positive time intervals, the tail decay rate, four relative amplitudes, four frequency parameters, and the merger frequency derivative. They are predicted by the network, while the branch flag is determined by the inputs. Positive parameters are represented logarithmically, and times and frequencies are made dimensionless using the total mass. During reference data construction, the initial frequency and its derivative are determined from the premerger waveform. During generation, the predicted parameters are substituted into the NRPM relations for piecewise amplitude, phase evolution, and exponential decay to recover the postmerger waveform. Branch selection remains determined by the tidal coupling parameter in Eq.~\eqref{eq:kappa}. We use dimensionless time $\tau=t/M_t$ and postmerger strain $\widehat h_{{\rm PM},+}(\tau)=(d_t/M_t)\widehat h_{{\rm PM},+}^{\rm phys}(M_t\tau)$, with $\mathrm{phys}$ denoting physical strain. Fourier transforming the 0.5\,s postmerger waveform with respect to $\tau$ and adding the premerger spectrum gives
\begin{equation}
\widehat{\mathcal H}_{\rm BNS}(u)=
\widehat{\mathcal H}_{\rm pre}(u)+
\mathcal F_\tau\!\left[\widehat h_{{\rm PM},+}(\tau>0)\right](u),
\label{eq:bns_full_fd}
\end{equation}
where $\mathcal F_\tau$ transforms dimensionless time to $u$, using the scaling of Eq.~\eqref{eq:dimensionless_waveform}. The premerger waveform includes the merger peak, so only $\tau>0$ is retained here. All three binary types use a common prediction vector with 304 components, comprising 288 spectral coefficients and the 16 dimensional postmerger representation. BBH and NSBH recover waveforms using only the spectral coefficients, whereas BNS also uses the physical postmerger decoder. All three output 4480 complex frequency samples.

\subsection{Shared network and modules for individual binary types}
The complete network architecture is shown in Fig.~\ref{fig:supp_architecture}.
The network first converts the 11 raw inputs into 22 dimensionless features, including $q$, $\eta$, mass asymmetry, the components of $\boldsymbol\chi_i$, $\Lambda_i$ and their combinations, and $n_i$. The effective aligned spin $\chi_{\rm eff}$~\cite{Ajith2011EffectiveSpin} and effective tidal deformability $\widetilde\Lambda$~\cite{Wade2014Tidal} are computed using Eq.~\eqref{eq:effective_physics} to introduce combinations of $\boldsymbol\chi_i$ and $\Lambda_i$ into the network,
\begin{equation}
\begin{split}
\chi_{\rm eff}&=\frac{q\chi_{1z}+\chi_{2z}}{1+q},\\
\widetilde\Lambda&=\frac{16}{13}
\frac{(q+12)q^4\Lambda_1+(1+12q)\Lambda_2}{(1+q)^5}.
\end{split}
\label{eq:effective_physics}
\end{equation}
Transformations such as $\log(1+\Lambda_i)$ are applied to $\Lambda_i$ to reduce their numerical range. The 22 dimensional feature vector $\boldsymbol v$ is then mapped to Fourier features using Eq.~\eqref{eq:fourier_features}, retaining the original features and adding four sets of sine and cosine components at different scales,
\begin{equation}
\gamma(\boldsymbol v)=\left[\boldsymbol v,
\{\sin(\pi2^k\boldsymbol v),\cos(\pi2^k\boldsymbol v)\}_{k=0}^{3}\right].
\label{eq:fourier_features}
\end{equation}
The resulting 198 features are linearly mapped to a 256 dimensional representation, added to a learnable embedding of $(n_1,n_2)$, and passed through six gated residual blocks. Each block uses layer normalization (LN), a twofold channel expansion, and sigmoid linear unit (SiLU) gating to update the input features $\boldsymbol z$ according to Eq.~\eqref{eq:gated_residual},
\begin{equation}
\begin{split}
\boldsymbol z' &= \boldsymbol z+\boldsymbol a\odot W_d
\left[\boldsymbol v_z\odot\operatorname{SiLU}(\boldsymbol g_z)\right],\\
(\boldsymbol v_z,\boldsymbol g_z)&=W_u\operatorname{LN}(\boldsymbol z),
\end{split}
\label{eq:gated_residual}
\end{equation}
where $W_u$ and $W_d$ are linear maps and $\operatorname{LN}$ denotes layer normalization. The features $\boldsymbol v_z$ provide the update, while $\boldsymbol g_z$ modulates its amplitude after SiLU activation. The result is added to the original input through a residual connection. The channel scale parameters $\boldsymbol a$ are initialized to 0.1.

The shared features are passed to a shared output layer and branches specific to the binary types, producing a base prediction and dedicated outputs. Each dedicated branch contains three residual blocks and a linear layer with 304 outputs. The binary type $s\in\{\mathrm{BBH},\mathrm{BNS},\mathrm{NSBH}\}$ is determined from $(n_1,n_2)$ to select the branch. For type $s$, $\overline c_{sj}=(c_{sj}-\mu_{sj})/\sigma_{sj}$ uses training means $\mu_{sj}$ and standard deviations $\sigma_{sj}$. Adding the shared and gated branch outputs gives the initial standardized prediction,
\begin{equation}
\widehat{\overline{\boldsymbol c}}^{(0)}_s=
H_{\rm common}(\boldsymbol z)+g_s(\boldsymbol z)H_s(\boldsymbol z).
\label{eq:shared_expert}
\end{equation}
Here $H_{\rm common}$ and $H_s$ denote the outputs of the shared output layer and the corresponding branch for the binary type. The gating weight $g_s$ is computed from the shared features using a sigmoid function and controls the contribution of the dedicated branch to the initial prediction. For BNS, a separate postmerger parameter prediction network concatenates the 256 shared features with the 198 Fourier features to give 454 inputs, maps them linearly to 256 dimensions, and passes them through four residual blocks. Its 16 dimensional output replaces the last 16 components of the initial prediction, with the branch flag set from the inputs after restoring coefficient units.

The BBH and NSBH reference waveforms contain higher order radiation modes, whose superposition produces more complex spectral amplitude and phase variations. To describe their phase structure in greater detail, we introduce spectral refinement networks after the initial coefficient prediction to correct the low and high frequency amplitude and phase coefficients separately. For each of BBH and NSBH, the refinement network concatenates the 198 Fourier features with the standardized prediction of 304 components to obtain 502 inputs. These are linearly mapped to 256 dimensions and passed through four residual blocks, followed by separate low and high frequency output branches. Each output branch contains two residual blocks and produces 96 or 192 coefficient corrections, respectively. The concatenated corrections are added to the first 288 spectral coefficients of the initial prediction, while the remaining 16 components are unchanged. The complete model has 14,730,515 parameters, comprising 7,998,675 in the shared network and 6,731,840 in the two spectral refinement networks. Initial predictions and refinement additions use standardized coordinates; decoding first restores coefficient units, $\widehat c_{sj}=\mu_{sj}+\sigma_{sj}\widehat{\overline c}_{sj}$.

\subsection{Staged training and loss functions}
\begin{figure*}[tbp]
\centering
\resizebox{\textwidth}{!}{%
\begin{tikzpicture}[
  x=1mm,y=1mm,
  font=\sffamily\fontsize{9.5}{11.5}\selectfont,
  line width=0.45pt,
  >={Stealth[length=1.8mm,width=1.2mm]},
  stage/.style={draw=black,fill=white,text=black,align=center,
    text width=36mm,minimum height=17mm,inner xsep=1mm,inner ysep=2mm,
    outer sep=0pt}
]
  \node[stage] (joint) at (0,0)
    {Joint training\\[2pt]8192 / type};
  \node[stage] (jointmore) at (46,0)
    {Joint continuation\\[2pt]$10^5$ / type};
  \node[stage] (specific) at (92,0)
    {Training specialized\\modules\\[2pt]$10^5$ / type};
  \node[stage] (continued) at (138,0)
    {Specialized training\\[2pt]$10^6$ / type};
  \draw[->] (joint.east) -- (jointmore.west);
  \draw[->] (jointmore.east) -- (specific.west);
  \draw[->] (specific.east) -- (continued.west);
\end{tikzpicture}%
}
\caption{\textbf{Staged training of URSA-GW.} The sample size per binary type increases from 8192 to $10^5$ and $10^6$. The 100,000 samples are first used for joint continuation, followed by specialized training with the shared components fixed. The million sample stage retains the same settings. BBH and NSBH optimize their spectral refinement networks, while BNS optimizes its dedicated branch and postmerger parameter network.}
\label{fig:training}
\end{figure*}
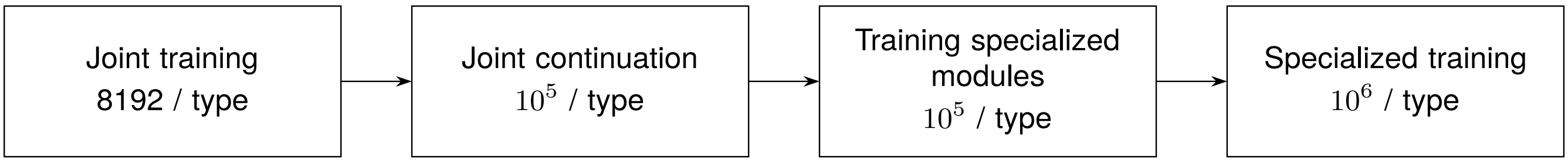

The staged training procedure is shown in Fig.~\ref{fig:training}. Training begins with joint learning across the three binary types using 8192 samples per type. The shared layers, shared output layer, gates, dedicated branches for all binary types, and BNS postmerger parameter network are updated together to learn the mapping from binary parameters to spectral coefficients and postmerger physical parameters. Each batch contains 512 samples from each binary type, and the three types contribute equally. Spectral compression bases, target coefficient normalization parameters, and loss scales are determined from this initial training set and remain fixed in later stages, keeping the waveform representation and training targets consistent as the dataset expands. In Eq.~\eqref{eq:coefficient_losses}, $c_j$ and $\widehat c_j$ are unstandardized target and predicted coefficients for binary type $s$. The weighted coefficient loss uses fixed scales $\alpha_{sj}$, while the standardized spectral loss divides by the training standard deviations $\sigma_{sj}$,
\begin{equation}
\begin{split}
\mathcal L_{\rm coef}&=\frac1{304}\sum_{j=1}^{304}
\alpha_{sj}^2(\widehat c_j-c_j)^2,\\
\mathcal L_{\rm coef,std}&=\frac1{288}\sum_{j=1}^{288}
\left[(\widehat c_j-c_j)/\sigma_{sj}\right]^2 .
\end{split}
\label{eq:coefficient_losses}
\end{equation}
For BNS, Eq.~\eqref{eq:pm_waveform_envelope} also constrains the complex postmerger waveform and its amplitude envelope, while Eq.~\eqref{eq:pm_losses} constrains the unwrapped phase and instantaneous frequency. Writing the reference waveform as $h=Ae^{i\phi}$ and the predicted waveform as $\widehat h=\widehat A e^{i\widehat\phi}$, the losses are
\begin{align}
\mathcal L_{\rm PM}&=\frac{\sum_t|\widehat h-h|^2}{\sum_t A^2},&
\mathcal L_A&=\frac{\sum_t(\widehat A-A)^2}{\sum_t A^2},\label{eq:pm_waveform_envelope}\\
\mathcal L_\phi&=\frac{\sum_t A^2(\widehat\phi-\phi)^2}{\sum_t A^2},&
\mathcal L_f&=\frac{\sum_t A^2(\widehat f-f)^2}{\sum_t A^2 f^2}.
\label{eq:pm_losses}
\end{align}
Early training first optimizes the coefficients, phase, frequency, and envelope to bring the predicted postmerger oscillations close to the reference, before introducing the complex waveform loss. After 10 epochs of warmup, the phase root mean square error $\sqrt{\mathcal L_\phi}$ is evaluated on BNS training diagnostic samples. When its median is below 0.1\,rad and its 90th percentile is below 0.3\,rad, the complex waveform loss weight is increased linearly to its prescribed value over 10 epochs, starting from the next epoch. After expansion to $10^5$ training samples per binary type, joint training continues with the same set of updated parameters, followed by specialized training for each binary type. For BBH and NSBH, all shared network parameters are fixed, and only the respective spectral refinement networks are updated. For BNS, only its dedicated branch, postmerger parameter prediction network, and associated input mapping are updated. The shared layers, shared output layer, gates, and branches for the other binary types remain fixed. Subsequent training therefore focuses on waveform details of each binary type while preserving the mapping learned by the shared network. When the training set is further expanded to $10^6$ samples per binary type, the preceding weights initialize continued training of the spectral refinement networks and BNS dedicated branch, with the same parameter freezing settings.

The BBH and NSBH spectral refinement networks constrain the generated complex frequency domain (FD) spectrum, logarithmic amplitude, and unwrapped phase jointly. The spectral representation uses a boundary at $u=0.012$ to define different sampling grids and PCA bases for compression in the two intervals. The training loss instead uses the high frequency error evaluation boundary at $u=0.02$, with separate normalization and equal weighting on either side. This prevents the larger squared amplitudes at low frequencies from dominating the total loss and also constrains amplitude and phase deviations at high frequencies. Thus, $u=0.012$ sets the boundary for spectral representation, and $u=0.02$ sets the boundary for band weighting in BBH and NSBH refinement training. Denoting the reference and generated complex spectra by $h$ and $\widehat h$, we define the logarithmic amplitude difference $\Delta a=\log|\widehat h|-\log|h|$ and unwrapped phase difference $\Delta\phi=\widehat\phi-\phi$. In each band $b$, the weights $w_k\propto|h_k|^2\Delta u_k$ are determined by the squared reference amplitude and frequency integration spacing and normalized so that $\sum_{k\in b}w_k=1$. Equation~\eqref{eq:spectral_complex_loss} measures relative error through the complex ratio of the generated and reference spectra, jointly constraining amplitude and phase agreement. Equation~\eqref{eq:spectral_shape_losses} separately constrains deviations in logarithmic amplitude and unwrapped phase,
\begin{align}
\mathcal L_{{\rm FD},b}&=\sum_{k\in b}w_k
\left|e^{\Delta a_k+i\Delta\phi_k}-1\right|^2,\label{eq:spectral_complex_loss}\\
\begin{split}
\mathcal L_{{\rm log}A,b}&=\sum_{k\in b}w_k(\Delta a_k)^2,\\
\mathcal L_{\phi,b}&=\sum_{k\in b}w_k(\Delta\phi_k)^2.
\end{split}
\label{eq:spectral_shape_losses}
\end{align}
Equation~\eqref{eq:corr_loss} combines the losses of the two bands with equal weights and adds the standardized mean squared error of the first 288 spectral coefficients, constraining both the waveforms and coefficients predicted by the refinement networks,
\begin{equation}
\begin{split}
\mathcal L_{\rm corr}={}&\frac12\sum_b\left[
\mathcal L_{{\rm FD},b}+0.25\mathcal L_{{\rm log}A,b}
+0.25\mathcal L_{\phi,b}\right]\\
&+0.01\mathcal L_{\rm coef,std},
\end{split}
\label{eq:corr_loss}
\end{equation}
During training, BBH and NSBH contribute equally to the loss. Shared network parameters remain fixed, and only the respective spectral refinement networks are updated.

BNS specialized training updates only the BNS dedicated branch, postmerger parameter prediction network, and associated input mapping, totaling 2,966,592 parameters. The shared layers, shared output layer, gates, and branches for other binary types remain fixed. The total loss in Eq.~\eqref{eq:bns_loss} jointly constrains coefficient predictions, the premerger spectrum, and the amplitude, phase, and frequency of the postmerger waveform,
\begin{equation}
\begin{split}
\mathcal L_{\rm BNS}={}&\mathcal L_{\rm coef}+3\mathcal L_{\rm PM}\\
&+0.3(\mathcal L_\phi+\mathcal L_f+\mathcal L_A)
+\frac12\sum_b\mathcal L_{{\rm pre},b},
\end{split}
\label{eq:bns_loss}
\end{equation}
where $\mathcal L_{{\rm pre},b}$ measures the relative complex spectral error before merger. The two bands follow the spectral representation boundary at $u=0.012$. The postmerger waveform is constrained jointly by the complex waveform loss $\mathcal L_{\rm PM}$ and the phase, instantaneous frequency, and amplitude envelope losses. To reduce training costs, frequency domain losses use every eighth frequency node, and postmerger time domain losses use every fourth time node. Evaluation uses all 4480 frequency nodes and 16385 postmerger time nodes.

All training stages use adaptive moment estimation with decoupled weight decay (AdamW), setting the weight decay to $10^{-6}$ and the gradient norm clipping threshold to 5. The initial joint training learning rate is $3\times10^{-4}$. After expansion to $10^5$ samples per binary type, the learning rate for joint continuation is reduced to $10^{-5}$, and the spectral refinement networks use $10^{-4}$. At the $10^6$ sample stage, both the spectral refinement networks and BNS dedicated branch continue training with a learning rate of $10^{-5}$. If the validation objective does not improve for five consecutive epochs, the learning rate is halved, down to a minimum of $10^{-7}$. Training stops after 30 consecutive epochs without improvement, and the weights with the best validation performance are restored. Training and validation use 32 bit floating point (FP32) precision. TensorFloat-32 (TF32) acceleration is allowed during initial training of the shared network and disabled for validation. TF32 is disabled for continued training of both the spectral refinement networks and the BNS dedicated branch at the million sample stage. Experiments use PyTorch 2.6.0 on an NVIDIA A800-SXM4-80GB.

\subsection{Waveform evaluation}
We evaluate agreement between generated and reference waveforms using waveform mismatch~\cite{Sathyaprakash2009,Lindblom2008}. Over all 4480 frequency nodes, we define the complex inner product $\langle a,b\rangle_w=\sum_k w_ka_k^*b_k$ and compute the mismatch using Eq.~\eqref{eq:mismatch},
\begin{equation}
\mathcal M=1-
\frac{|\langle h_{\rm ref},h_{\rm AI}\rangle_w|}
{\sqrt{\langle h_{\rm ref},h_{\rm ref}\rangle_w
\langle h_{\rm AI},h_{\rm AI}\rangle_w}},
\label{eq:mismatch}
\end{equation}
where $h_{\rm ref}$ and $h_{\rm AI}$ denote the complex reference spectrum and the complex spectrum generated by artificial intelligence (AI), respectively. Weights $w_k$ are determined by the integration spacing of the nonuniform frequency grid, without detector noise weighting. Endpoint weights are half the adjacent frequency spacing, and interior weights are $w_k=(u_{k+1}-u_{k-1})/2$. Taking the absolute value of the complex inner product removes an overall constant phase difference between the waveforms. Reference and generated waveforms share the same time origin, with no additional search over time shifts. Smaller mismatch indicates closer agreement. Frequency domain plots directly use the amplitudes of the reference and model decoded spectra.

To display waveforms in the time domain, we first remove the leading Newtonian phase from the spectrum, interpolate the residual spectrum onto a uniform frequency grid, and restore the leading phase and physical units. A smooth 10\% taper is applied at low frequencies, followed by an inverse Fourier transform. Both reference and generated waveforms use the reference peak time as their common time origin, with the same time axis and display window. The generated waveform is not independently aligned to its peak. Both waveforms are normalized by the maximum absolute strain of the complete reference record, $h_{\rm pk}=\max_t|h_{\rm ref}(t)|$. The complete waveform and enlarged views for each sample use this same normalization, with enlarged views changing only the displayed range.

We calculate waveform cycles and durations from the first $10^5$ training reference waveforms per binary type. Cycle counts are obtained from the complete $h_+(t)$ recovered from the original reference spectrum, using the Hilbert transform and phase unwrapping as in the published conditional autoencoder (cAE) study~\cite{Sun2025cAE} and Eq.~\eqref{eq:cycles},
\begin{equation}
\begin{split}
s_a(t)&=h_+(t)+i\mathsf H[h_+(t)],\\
\phi_a(t)&=\operatorname{unwrap}\arg s_a(t),\\
N_{\rm cyc}&=\frac{\phi_a(t_{\rm end})-\phi_a(t_{\rm start})}{2\pi},
\end{split}
\label{eq:cycles}
\end{equation}
where $\mathsf H$ is the Hilbert transform and $\phi_a$ is the unwrapped analytic phase. The difference between its final and initial values divided by $2\pi$ gives the number of waveform cycles. Only exactly zero samples at the two ends of the record are removed, and BNS includes the 0.5\,s postmerger segment. Duration is the length of the same complete band limited time window, $\Delta t=t_{\rm end}-t_{\rm start}$. Duration ranges are listed in Table~\ref{tab:domain}. For BBH at fixed $q$, $\boldsymbol\chi_i$, and dimensionless frequency band, waveform duration scales linearly with total mass,
\begin{equation}
\Delta t(M)=\Delta t(M_{\rm ref})\frac{M}{M_{\rm ref}},\qquad M_{\rm ref}=60M_\odot.
\label{eq:duration_mass_scaling}
\end{equation}
Equation~\eqref{eq:duration_mass_scaling} gives duration ranges for different total mass intervals. At the reference mass, BBH durations are 110.19--360.70\,s. For a total mass interval $[M_{\rm lo},M_{\rm hi}]$, the lower and upper duration bounds follow by multiplying the shortest reference duration by $M_{\rm lo}/M_{\rm ref}$ and the longest reference duration by $M_{\rm hi}/M_{\rm ref}$. Mass scaling preserves the number of waveform cycles, while physical frequencies scale inversely with total mass. Duration here describes the physical time covered by a waveform and differs from the computation time required to generate it. Examples in the main text are selected from the test set, requiring small full band mismatch and cycle counts between the 25th and 75th percentiles of the corresponding training reference distribution, with further selection based on agreement in high frequency spectra and waveforms near or after merger. Overall accuracy is assessed using all test samples.

\subsection{Computational optimization and efficiency tests}
Different waveform decoding operations are assigned to the CPU or GPU according to binary type. Reuse of precomputed quantities and batch operations improves the efficiency of converting network predictions into physical waveforms. Spectral bases, frequency factors of the analytic baseline, and interpolation weights are precomputed and cached during initialization to avoid repeated calculations. For BBH and NSBH, the CPU performs the linear projection of spectral coefficients, while the GPU restores amplitude and phase and assembles complex spectra in parallel. For BNS, spectral reconstruction, physical postmerger decoding, the fast Fourier transform (FFT), and frequency interpolation are all performed on the GPU, reducing data transfer between devices. BNS decoding uses double precision and retains 0.5\,s of postmerger evolution. Zero padding to a 4\,s window is used for a Fourier transform with 131072 points. All three binary types output complex frequency domain waveforms with 4480 points and support batch generation grouped by binary type. Across 4096 test waveforms per binary type, the full band mismatch between the optimized and original implementations never exceeds $4.1\times10^{-11}$, showing that the computational acceleration preserves numerical waveform agreement.

Computational efficiency is evaluated at batch sizes of 1, 256, 512, 1024, and 2048. Reference models run in parallel on 64 physical cores of an Intel Xeon Platinum 8480+ server, while URSA-GW uses an NVIDIA A800-SXM4-80GB. Both use the same validation parameters, and timing covers the complete procedure from intrinsic parameter input to the output of 4480 complex frequency samples.

\section*{Acknowledgments}
\begingroup
\raggedright
\noindent\textbf{Funding}\quad This work was supported by the National Key Research and Development Program of China (2023YFC2206702), the Fundamental Research Funds for the Central Universities (2024IAIS-ZD009), and the National Natural Science Foundation of China (12575072, 12347101).
\endgroup

\bibliographystyle{unsrtnat}
\bibliography{references}
\clearpage
\onecolumn
\setcounter{figure}{0}
\renewcommand{\thefigure}{S\arabic{figure}}
\renewcommand{\theHfigure}{supp.\arabic{figure}}
\pdfbookmark[0]{Supplementary Materials}{supplementary}
\section*{Supplementary Materials}
\captionsetup[figure]{name={Fig.},font=footnotesize}
\begin{figure}[htbp]
\centering
\resizebox{0.90\textwidth}{!}{
\begingroup
\definecolor{ArchBBH}{HTML}{C58B25}
\definecolor{ArchBNS}{HTML}{3678B5}
\definecolor{ArchNSBH}{HTML}{168C87}
\definecolor{ArchInk}{HTML}{142B43}
\newcommand{\adetail}[1]{{\fontsize{8.5}{10.2}\selectfont #1}}
\hyphenpenalty=10000\exhyphenpenalty=10000
\begin{tikzpicture}[x=1cm,y=1cm,text=ArchInk,
 font=\sffamily\fontsize{9.5}{11.2}\selectfont,line cap=round,line join=round,
 box/.style={draw=ArchInk,line width=.7pt,rounded corners=.7mm,fill=white,align=center,minimum height=.8cm,inner sep=4pt},
 module/.style={box,line width=1pt},
 panel/.style={draw=ArchBNS!80,dashed,line width=.75pt,rounded corners=1.5mm},
 title/.style={anchor=west,font=\sffamily\bfseries\fontsize{12.5}{14}\selectfont,fill=ArchBNS!10,inner xsep=5pt,inner ysep=3pt},
 small/.style={font=\sffamily\fontsize{8.5}{10.2}\selectfont,align=center},
 flow/.style={-{Stealth[length=1.8mm,width=1.2mm]},draw=black,line width=.75pt},
 wire/.style={draw=black,line width=.75pt},
 op/.style={circle,draw=black,fill=white,line width=.75pt,minimum size=.55cm,inner sep=0pt,font=\fontsize{13}{14}\selectfont}]
\path[use as bounding box] (0,0) rectangle (22,24.4);
\node[font=\sffamily\bfseries\fontsize{16}{19}\selectfont] at (11,24.05) {URSA-GW: Detailed architecture};

\draw[panel] (0,18.4) rectangle (22,23.6);
\node[title] at (.25,23.16) {(a)\quad From binary parameters to waveforms};
\node[module,text width=2.3cm,minimum height=1.4cm] (input) at (1.65,20.9)
 {$\boldsymbol x$\\[3pt]$q$\\$\boldsymbol\chi_1,\boldsymbol\chi_2$\\$\Lambda_1,\Lambda_2$\\$(n_1,n_2)$};
\node[module,text width=2.65cm,minimum height=1.4cm,fill=ArchBNS!5] (shared) at (5.2,20.9)
 {\textbf{Shared neural network (b)}\\[3pt]\adetail{Initial coefficients\\Shared features\\Fourier features}};
\node[module,draw=ArchBBH,text width=3.0cm] (bbhc) at (10,22.15) {\textbf{BBH refinement (c)}};
\node[box,text width=2.65cm] (bbhd) at (14.6,22.15) {Spectral decoder};
\node[module,draw=ArchBBH,text width=2.6cm] (bbho) at (19.55,22.15) {\textbf{BBH waveform}};
\node[module,draw=ArchBNS,text width=3.0cm] (bnsd) at (12.3,20.9) {\textbf{BNS decoder (d)}};
\node[module,draw=ArchBNS,text width=2.6cm] (bnso) at (19.55,20.9) {\textbf{BNS waveform}};
\node[module,draw=ArchNSBH,text width=3.0cm] (nsbhc) at (10,19.65) {\textbf{NSBH refinement (c)}};
\node[box,text width=2.65cm] (nsbhd) at (14.6,19.65) {Spectral decoder};
\node[module,draw=ArchNSBH,text width=2.6cm] (nsbho) at (19.55,19.65) {\textbf{NSBH waveform}};
\draw[flow] (input)--(shared);
\draw[wire] (shared.east)--(7.35,20.9);
\draw[flow] (7.35,20.9)|-(bbhc.west);
\draw[flow] (7.35,20.9)--(bnsd.west);
\draw[flow] (7.35,20.9)|-(nsbhc.west);
\draw[flow] (bbhc)--(bbhd); \draw[flow] (bbhd)--(bbho);
\draw[flow] (bnsd)--(bnso);
\draw[flow] (nsbhc)--(nsbhd); \draw[flow] (nsbhd)--(nsbho);
\node[small] at (14.6,18.88) {Spectral decoder: PCA reconstruction $\times$ analytic baseline};

\draw[panel] (0,11.9) rectangle (22,18.05);
\node[title] at (.25,17.61) {(b)\quad Shared neural network};
\node[box,text width=1.6cm] (physics) at (1.15,15.35) {Physics\\features\\\adetail{11 $\to$ 22}};
\node[box,text width=1.65cm] (fourier) at (3.6,15.35) {Fourier\\mapping\\\adetail{198 features}};
\node[box,text width=1.25cm] (dense) at (5.83,15.35) {Dense\\\adetail{256}};
\node[op] (embedsum) at (7.3,15.35) {$+$};
\node[module,text width=2.05cm] (trunk) at (9.18,15.35) {Shared layers\\\adetail{Block (e) $\times\,6$}};
\node[box,text width=1.6cm] (features) at (11.56,15.35) {Shared\\features\\\adetail{256}};
\node[box,text width=2.25cm] (embedding) at (5.45,13.15) {$(n_1,n_2)$\\embedding};
\node[box,text width=3.1cm] (common) at (15.05,16.9) {Shared output layer\\\adetail{LN $\to$ Dense 304}};
\node[box,text width=3.1cm] (expert) at (15.05,15.18) {Binary-specific network\\\adetail{Block (e) $\times\,3$\\LN $\to$ Dense 304}};
\node[box,text width=3.1cm] (gate) at (15.05,13.56) {Gate network\\\adetail{LN $\to$ Dense 3\\$\to$ Sigmoid}};
\node[op] (weight) at (17.85,15.18) {$\times$};
\node[op] (combine) at (18.85,16.9) {$+$};
\node[module,text width=1.95cm] (coeff) at (20.58,16.9) {Initial\\coefficients\\\adetail{304}};
\node[small,text width=2.65cm] at (20.25,14.35) {288 spectral coefficients\\+ 16 BNS representation entries};
\draw[flow] (physics)--(fourier); \draw[flow] (fourier)--(dense);
\draw[flow] (dense)--(embedsum); \draw[flow] (embedsum)--(trunk);
\draw[flow] (trunk)--(features);
\draw[flow] (embedding.east)-|(embedsum.south);
\draw[wire] (features.east)--(12.83,15.35);
\draw[flow] (12.83,15.35)|-(common.west);
\draw[flow] (12.83,15.35)|-(expert.west);
\draw[flow] (12.83,15.35)|-(gate.west);
\draw[flow] (expert)--(weight);
\draw[flow] (gate.east)-|(weight.south);
\draw[flow] (weight.east)-|(combine.south);
\draw[flow] (common)--(combine); \draw[flow] (combine)--(coeff);
\node[small] at (6.05,12.28) {$\widehat{\overline{\boldsymbol c}}^{(0)}_s=H_{\rm common}(\boldsymbol z)+g_s(\boldsymbol z)H_s(\boldsymbol z)$};
\node[small] at (15.4,12.45) {Select $H_s$ and $g_s$ using $(n_1,n_2)$};

\draw[panel] (0,5.1) rectangle (10.7,11.55);
\node[title] at (.25,11.1) {(c)\quad Spectral refinement network};
\node[box,text width=2.5cm] (corrin) at (1.5,10) {Coefficients +\\Fourier features\\\adetail{304 + 198}};
\node[box,text width=1.25cm] (corrdense) at (4.3,10) {Dense\\\adetail{256}};
\node[module,text width=2.65cm] (corrtrunk) at (7.6,10) {Shared layers\\\adetail{Block (e) $\times\,4$}};
\node[box,text width=3.0cm] (low) at (4.1,8.2) {Low-frequency branch\\\adetail{Block (e) $\times\,2$\\LN $\to$ Dense 96}};
\node[box,text width=3.0cm] (high) at (8.0,8.2) {High-frequency branch\\\adetail{Block (e) $\times\,2$\\LN $\to$ Dense 192}};
\node[box,text width=2.2cm,minimum height=.6cm] (concat) at (6.05,6.83) {Concatenate\\\adetail{288}};
\node[box,text width=2.65cm,minimum height=.7cm] (initial) at (1.85,5.8) {Initial spectral\\coefficients};
\node[op] (corrsum) at (6.05,5.8) {$+$};
\node[module,text width=2.7cm,minimum height=.7cm] (corrected) at (8.9,5.8) {Refined spectral\\coefficients};
\draw[flow] (corrin)--(corrdense); \draw[flow] (corrdense)--(corrtrunk);
\draw[wire] (corrtrunk.south)--(7.6,9.25);
\draw[flow] (7.6,9.25)-|(low.north); \draw[flow] (7.6,9.25)-|(high.north);
\draw[flow] (low.south)|-(concat.west); \draw[flow] (high.south)|-(concat.east);
\draw[flow] (concat)--(corrsum); \draw[flow] (initial)--(corrsum);
\draw[flow] (corrsum)--(corrected);

\draw[panel] (11.0,5.1) rectangle (22,11.55);
\node[title] at (11.25,11.1) {(d)\quad BNS decoder};
\node[box,text width=2.15cm] (bnscoeff) at (12.5,10.0) {Initial spectral\\coefficients};
\node[box,text width=3.0cm] (predecoder) at (15.7,10.0) {Spectral decoder};
\node[box,text width=2.2cm] (prespectrum) at (19.1,10.0) {Premerger\\spectrum};
\node[box,text width=2.15cm] (pmin) at (12.5,7.95) {Shared + Fourier\\features\\\adetail{256 + 198}};
\node[module,text width=2.9cm] (pmhead) at (15.7,7.95) {PM parameter network\\\adetail{Dense 256\\Block (e) $\times\,4$\\LN $\to$ Dense 16}};
\node[box,text width=1.4cm] (nrpm) at (19.3,7.95) {NRPM};
\node[box,text width=1.4cm,minimum height=.65cm] (fft) at (19.3,6.65) {FFT};
\node[op] (bnssum) at (21.2,6.65) {$+$};
\node[module,text width=2.65cm,minimum height=.65cm] (bnswave) at (19.3,5.65) {BNS waveform};
\draw[flow] (bnscoeff)--(predecoder); \draw[flow] (predecoder)--(prespectrum);
\draw[flow] (prespectrum.east)--(21.2,10.0)--(bnssum.north);
\draw[flow] (pmin)--(pmhead); \draw[flow] (pmhead)--(nrpm);
\draw[flow] (pmhead.north)--node[right=2pt,small] {$\widehat A_{\rm pk}$} (predecoder.south);
\draw[flow] (nrpm)--(fft); \draw[flow] (fft)--(bnssum);
\draw[flow] (bnssum.south)|-(bnswave.east);
\node[small,text width=3.6cm] at (14.1,6.1) {PM: postmerger\\This network replaces\\the initial 16 BNS outputs};

\draw[panel] (0,0) rectangle (22,4.75);
\node[title] at (.25,4.3) {(e)\quad Gated residual block};
\node[box,text width=1.35cm] (gx) at (1.1,2.45) {Input\\\adetail{256}};
\node[box,text width=2.0cm] (gln) at (3.65,2.45) {LayerNorm};
\node[box,text width=1.4cm] (gdense) at (6.3,2.45) {Dense\\\adetail{1024}};
\node[box,text width=1.65cm] (gvalue) at (9.05,3.16) {Value\\\adetail{512}};
\node[box,text width=1.65cm] (ggate) at (9.05,1.74) {Gate: SiLU\\\adetail{512}};
\node[op] (gprod) at (11.4,2.45) {$\odot$};
\node[box,text width=1.4cm] (gdout) at (13.55,2.45) {Dense\\\adetail{256}};
\node[box,text width=1.7cm] (gscale) at (16.05,2.45) {Channel\\scale};
\node[op] (gsum) at (18.15,2.45) {$+$};
\node[box,text width=1.6cm] (gout) at (20.25,2.45) {Output\\\adetail{256}};
\draw[flow] (gx)--(gln); \draw[flow] (gln)--(gdense);
\draw[wire] (gdense.east)--(7.55,2.45);
\draw[flow] (7.55,2.45)|-(gvalue.west); \draw[flow] (7.55,2.45)|-(ggate.west);
\draw[flow] (gvalue.east)-|(gprod.north); \draw[flow] (ggate.east)-|(gprod.south);
\draw[flow] (gprod)--(gdout); \draw[flow] (gdout)--(gscale);
\draw[flow] (gscale)--(gsum); \draw[flow] (gsum)--(gout);
\draw[flow] (gx.south)--(1.1,.7)--(18.15,.7)--(gsum.south);
\end{tikzpicture}
\endgroup}
\caption{\textbf{Detailed architecture of the Unified Residual Spectral Approximant for Gravitational Waves (URSA-GW).} \textbf{a,} The complete generation procedure from binary parameters to waveforms for the three binary types. \textbf{b,} The shared neural network combines the prediction of the shared output layer with the weighted output of the branch selected for that binary type to obtain the initial standardized coefficients $\widehat{\overline{\boldsymbol c}}^{(0)}_s$. Here $\boldsymbol z$ denotes shared features, $\boldsymbol v$ the physical input features, and $\gamma(\boldsymbol v)$ their Fourier feature mapping. $H_{\rm common}$ and $H_s$ denote the shared output mapping and the dedicated mapping for binary type $s$, respectively. $g_s$ is the gating weight for that binary type, and $s$ is determined by the binary type code $(n_1,n_2)$. \textbf{c,} Separate spectral refinement networks for binary black hole (BBH) and neutron star black hole (NSBH) waveforms each contain parallel low and high frequency branches. In standardized coordinates, corrections are added to the initial 288 spectral coefficients according to $\widehat{\overline{\boldsymbol c}}_s=\widehat{\overline{\boldsymbol c}}^{(0)}_{s,\rm spec}+\Delta\widehat{\overline{\boldsymbol c}}_s$. Coefficient units are restored before spectral decoding. \textbf{d,} For binary neutron star (BNS) waveforms, the postmerger (PM) parameter network predicts 15 continuous physical parameters, including the peak amplitude $\widehat A_{\rm pk}$ shared by both parts. One branch flag determined by the inputs completes the 16 dimensional representation. The numerical relativity postmerger (NRPM) model uses these parameters to generate the postmerger waveform in the time domain. A fast Fourier transform (FFT) converts it to a spectrum, which is added to the premerger spectrum rescaled by the same peak amplitude. \textbf{e,} The gated residual block used by the networks. Small text inside modules indicates layer structures and dimensions. LN denotes layer normalization, SiLU the sigmoid linear unit, and PCA principal component analysis. All three binary types output 4480 complex frequency samples. The network architecture and training methods are described in Materials and Methods.}
\label{fig:supp_architecture}
\end{figure}
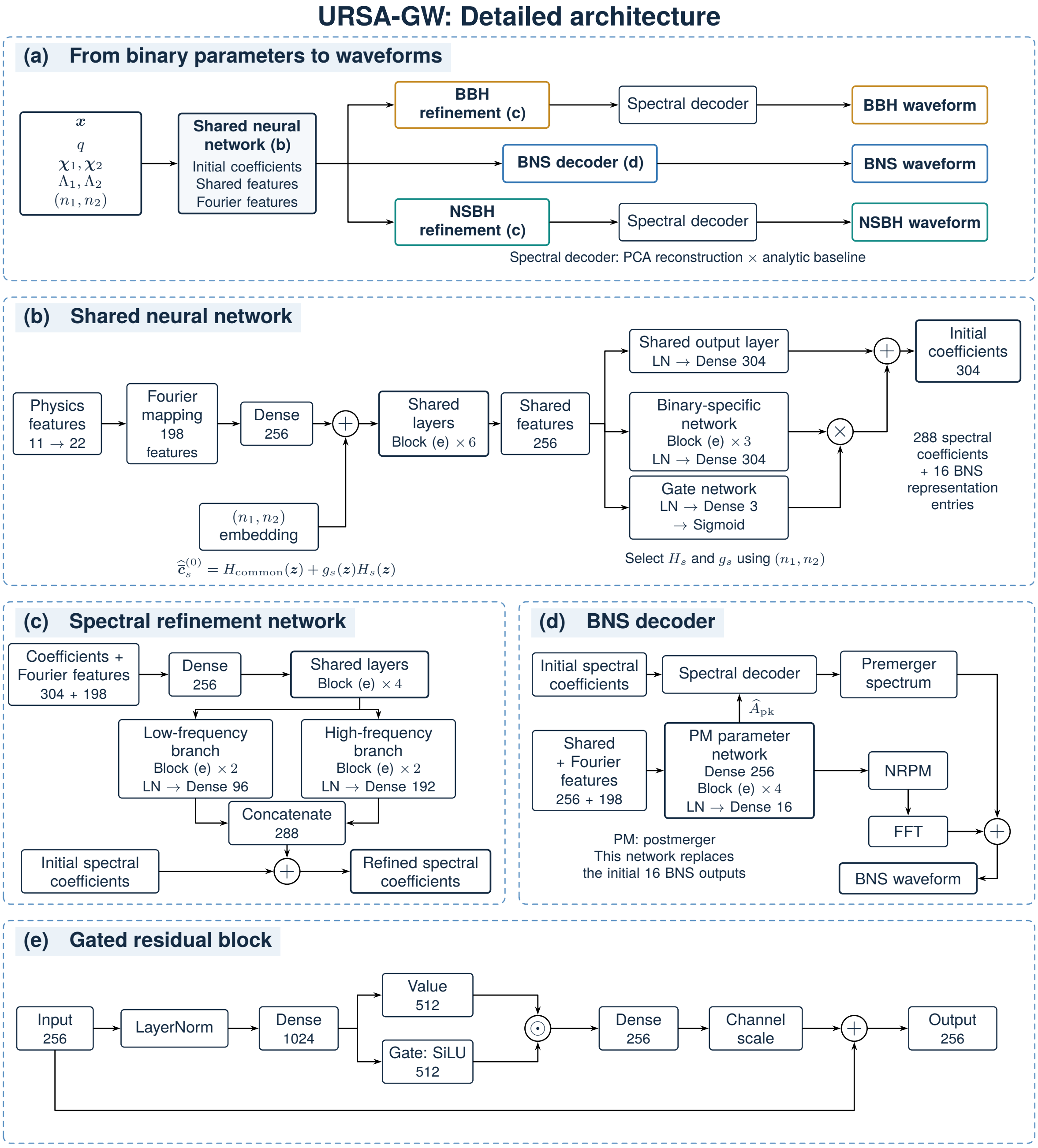

\end{document}